\documentclass[epsfig,pre,aps]{revtex4}
\usepackage{amsmath}
\usepackage{graphicx}
\usepackage{dcolumn}
\usepackage{bm}
\usepackage{subfigure}
\begin{document}

\title{Solitons in one-dimensional photonic crystals }
\author{\ Thawatchai Mayteevarunyoo$^{1,2}$ and B. A. Malomed$^{1}$}
\affiliation{$^1$Department of Physical Electronics, School of
Electrical Engineering,
\\Faculty of Engineering, Tel-Aviv
University, Tel-Aviv 69978, Israel} \affiliation{$^2$Department of
Telecommunication Engineering,
\\Mahanakorn University of Technology, Bangkok 10530, Thailand}

\begin{abstract}
We report results of a systematic analysis of spatial solitons in the model
of 1D photonic crystals, built as a periodic lattice of waveguiding
channels, of width $D$, separated by empty channels of width $L-D$. The
system is characterized by its structural \textquotedblleft duty cycle", $%
\mathrm{DC~}\equiv D/L$. In the case of the self-defocusing (SDF) intrinsic
nonlinearity in the channels, one can predict new effects caused by
competition between the linear trapping potential and the effective
nonlinear repulsive one. Several species of solitons are found in the first
two finite bandgaps of the SDF model, as well as a family of fundamental
solitons in the semi-infinite gap of the system with the self-focusing
nonlinearity. At moderate values of $\mathrm{DC}$ (such as $0.50$), both
fundamental and higher-order solitons$\mathrm{~}$ populating the second
bandgap of the SDF model suffer destabilization with the increase of the
total power. Passing the destabilization point, the solitons assume a
flat-top shape, while the shape of unstable solitons gets inverted, with
local maxima appearing in empty layers. In the model with narrow channels
(around $\mathrm{DC}=0.25$), fundamental and higher-order solitons exist
only in the first finite bandgap, where they are stable, despite the fact
that they also feature the inverted shape.
\end{abstract}

\maketitle

\section{Introduction and the model}

The potential offered by photonic crystals (PhCs) for various applications,
as well for the development of fundamental studies is well known \cite{Yabl1}%
-\cite{Mingaleev}. The combination of PhC waveguiding structures with the
Kerr or saturable nonlinearity may give rise to spatial solitons, that were
studied theoretically in various settings \cite{John}-\cite{switch}.
Experimentally, spatial solitons similar to those expected in PhCs have been
created in photorefractive media with optically induced lattices, using the
methods developed in Refs. \cite{Moti-theory}-\cite{Moti2}. In addition to
spatial solitons, possibilities of the creation of temporal solitons were
also considered in PhC and related nonlinear media \cite{Sipe}-\cite%
{Mantsyzov} (the temporal solitons in PhCs may resemble their counterparts
studied theoretically and experimentally in photonic-crystal fibers, see,
e.g., works. \cite{Herrmann}-\cite{Skryabin}).

The fundamental model of effectively one-dimensional (1D) PhCs is built as a
periodic array of material stripes, with linear refractive index $\left(
n_{0}\right) _{\mathrm{mat}}>1$ and Kerr coefficient $\left( n_{2}\right) _{%
\mathrm{mat}}\neq 0$, alternating with empty layers (filled with air), that
have $n_{0}=1$ and $n_{2}=0$. In a more general case, one may consider an
``all-solid" model, which assumes an alternation of stripes made of
different materials. The structure based on the periodic alternation of
stripes with different local properties is usually called the Kronig-Penney
(KP) model \cite{KP}.

In the case when only $n_{0}$ is modulated according to the KP model, while $%
n_{2}\neq 0$ is constant, the model gives rise to 1D solitons and other
nonlinear states that were studied in Refs. \cite{Smerzi,Carr}. A
combination of the linear KP\ part with the uniform cubic-quintic
nonlinearity was studied too \cite{Gisin,China}. A characteristic feature of
the latter model is bistability of solitons and a plethora of stable
multi-peak localized patterns.

In a realistic PhC model, both the linear and nonlinear local coefficients
should be periodically modulated . The corresponding nonlinear Schr\"{o}%
dinger (NLS) equation for the spatial evolution of the complex amplitude of
the electromagnetic field, $\Psi (x,z)$, along propagation distance $z$
takes the following normalized form:

\begin{equation}
i\Psi _{z}+\Psi _{xx}+W(x)\left( 1+\sigma \left\vert \Psi \right\vert
^{2}\right) \Psi .  \label{NLS}
\end{equation}%
Here, term $\Psi _{xx}$ accounts for the transverse diffraction in the usual
paraxial approximation ($x$ is the transverse coordinate), and the KP
modulation function describes a periodic array of guiding channels, each of
width $D$ and depth $U>0$, which are separated by buffer (empty) layers of
width $L-D$, i.e., $D/L\equiv \mathrm{DC}$ may be considered as the ``duty
cycle" of the underlying material pattern:%
\begin{equation}
W(x)=\left\{
\begin{array}{cc}
0, & D+Ln<x<L\left( 1+n\right) -D \\
U, & Ln<x<D+Ln%
\end{array}%
\right. ,~n=0,\pm 1,\pm 2...~.  \label{W}
\end{equation}%
Coefficient $\sigma =\pm 1$ in Eq. (\ref{NLS}) determines the sign of the
nonlinearity, $\sigma =+1$ and $-1$ corresponding, severally, to the
self-focusing (SF) and self-defocusing (SDF)\ material. By means of an
obvious rescaling, one can fix $L\equiv 2\pi $ [and also $|\sigma |\equiv 1$%
, leaving the total power, $Q=\int_{-\infty }^{+\infty }\left\vert \Psi
(x)\right\vert ^{2}dx$, as a control parameter of stationary solutions].

The limit case corresponding to $\mathrm{DC}\rightarrow 0,~U\rightarrow
\infty $ is the ``Dirac comb", with $W(x)$ replaced by a chain of $\delta $%
-functions. This variant of the PhC model was studied in Refs. \cite%
{Sukhorukov,Sukhorukov2}. Our objective is to consider the regular version
of the KP system, with an intention to explore effects of the $\mathrm{DC}$
parameter on properties of solitons. We report systematic results for fixed
depth $U=3$ (which adequately represents the generic case), and three
characteristic values of $\mathrm{DC}$, \textit{viz}., $0.75,~0.50$, and $%
0.25$, the second case being the most interesting one. The results (the
shape of the solitons, their stability, etc.) are quite different from those
reported in Refs. \cite{Sukhorukov,Sukhorukov2}. We are chiefly dealing with
the SDF nonlinearity, $\sigma =-1$, which is most promising for finding new
properties of solitons. Indeed, while the linear part of the NLS equation
tends to trap them in waveguiding channels, the repulsive nonlinearity
produces a \textit{competing effect}, tending to push the field into buffer
layers. A noteworthy consequences of the competition is the existence of a
nontrivial instability border for solitons in the second bandgap, see below.

In the case of the uniform SDF nonlinearity, which corresponds to Eq. (\ref%
{NLS}) with the last term replaced by $\left[ W(x)-|\Psi |^{2}\right] \Psi $%
, the periodic potential gives rise to gap solitons (GSs), which were
studied in detail in terms of Bose-Einstein condensates (BECs) \cite{Salerno}%
-\cite{Louis}. Solitons found in the present work for $\mathrm{DC}=0.75$ are
similar to the ordinary GSs, while those found for $\mathrm{DC}=0.50$ and $%
0.25$ are quite different from them. In terms of BEC, it is also possible to
consider solitons trapped in a purely nonlinear lattice, which corresponds
to Eq. (\ref{NLS}) with the last term replaced by $W(x)\left\vert \Psi
\right\vert ^{2}\Psi $, where modulation function $W(x)$ is sinusoidal \cite%
{Sakaguchi,Fibich}. In the latter context, the competition between linear
and nonlinear lattices was considered too \cite%
{Fatkhulla,lin-nonlin-competition}. However, our results are different from
those reported for the nonlinear and combined lattice potentials in the BEC
models, because in the PhC model the local nonlinearity does not change its
sign.

The paper is structured as follows. A brief account of the methods used for
the analysis is given in Section II; the linear spectrum of the model is
also presented in this section. Basic results for the SDF model are reported
in Section III. These include families of fundamental and multi-peak GSs in
the first two finite (alias Bragg-reflection) bandgaps, which are always
stable in the first bandgap, but may feature an instability threshold in the
second. The stability of the GSs is identified through the computation of
eigenvalues for modes of small perturbations, and is verified by direct
simulations. Also reported are (weakly unstable) antisymmetric \textit{%
subfundamental} solitons (named as per Ref. \cite{we}), which are found in
the second bandgap, and (generally, stable) antisymmetric bound states of
fundamental solitons (alias ``twisted modes", as per Ref. \cite{Kivshar}).
In Section IV, we present a summary of results for the SF version of the
model, i.e., the one with $\sigma =+1$ in Eq. (\ref{NLS}). These results are
not drastically different from those reported earlier in other models
combining the periodic potential and attractive Kerr nonlinearity \cite%
{Wang,Salerno} [in particular, the stability of soliton families found in
the semi-infinite (alias total-internal-reflection) gap obeys the known
Vakhitov-Kolokolov (VK) criterion \cite{VK}]. The paper is concluded by
Section V.

\section{The methods}

Stationary soliton solutions to Eq. (\ref{NLS}) are sought for in the
ordinary form, $\Psi (x,z)=e^{ikz}\Phi (x)$, where $k$ is a real propagation
constant, and real function $\Phi (x)$ obeys equation%
\begin{equation}
-k\Phi +\Phi _{xx}+W(x)\left( \Phi +\sigma \Phi ^{3}\right) =0.
\label{Solution}
\end{equation}%
Equation (\ref{Solution}) was solved numerically by means of the iterative
Newton's method. For finding spatially even solutions, the initial guess was
taken as $\Phi _{0}(x)=A_{0}\mathrm{sech}\left( a_{0}x\right) $, with
constants $A_{0}$ and $a_{0}$. Odd solutions (the above-mentioned
subfundamental and twisted modes) were generated by a different initial
guess, $\Phi _{0}(x)=A_{0}\sin \left( q_{0}x\right) \cdot \mathrm{sech}%
\left( a_{0}x\right) $, with another constant $q_{0}$. Various families of
localized solutions are characterized by the total power as a function of $k$%
, $Q=Q(k)$.

The stability of the stationary solutions has been identified through
eigenvalues of small-perturbation modes. To this end, the perturbed solution
was substituted in Eq. (\ref{NLS}) as $\Psi \left( x,z\right) =e^{ikz}\left[
\Phi (x)+u\left( x\right) e^{i\lambda z}+v^{\ast }\left( x\right)
e^{i\lambda ^{\ast }z}\right] $, where $\Phi _{0}(x)$ is a solution of Eq. (%
\ref{Solution}), while $u(x)$ and $v(x)$ are complex eigenmodes of the
infinitesimal perturbation that can grow exponentially at (generally,
complex) rate $\lambda $. The linearization gives rise to the eigenvalue
problem in the following form:%
\begin{equation}
\left(
\begin{array}{cc}
-\left( \frac{d^{2}}{dx^{2}}-k\right) -W\left( x\right) \left( 1+2\sigma
\Phi _{0}^{2}\right) & -\sigma W\left( x\right) \Phi _{0}^{2} \\
\sigma W\left( x\right) \Phi _{0}^{2} & \left( \frac{d^{2}}{dx^{2}}-k\right)
+W\left( x\right) \left( 1+2\sigma \Phi _{0}^{2}\right)%
\end{array}%
\right) \left(
\begin{array}{c}
u \\
v%
\end{array}%
\right) =\lambda \left(
\begin{array}{c}
u \\
v%
\end{array}%
\right) ,  \label{lambda}
\end{equation}%
the underlying solution $\Phi _{0}$ being stable if all the eigenvalues are
real. Equations (\ref{lambda}) have been solved numerically with the help of
a finite-difference method based on Taylor-series expansions.

Before proceeding to the presentation of results for soliton families, it is
necessary to display the bandgap spectrum of the linearized version of Eq. (%
\ref{Solution}), as the location of solitons should be identified with
respect to this spectrum. It has been computed by means of software package
\texttt{SpectrUW }\cite{spectrum}. Characteristic examples of the spectrum
are displayed in Fig. 1 for three different values of the $\mathrm{DC}$.

\begin{figure}[h]
\centering\subfigure[]{\includegraphics[width=3in]{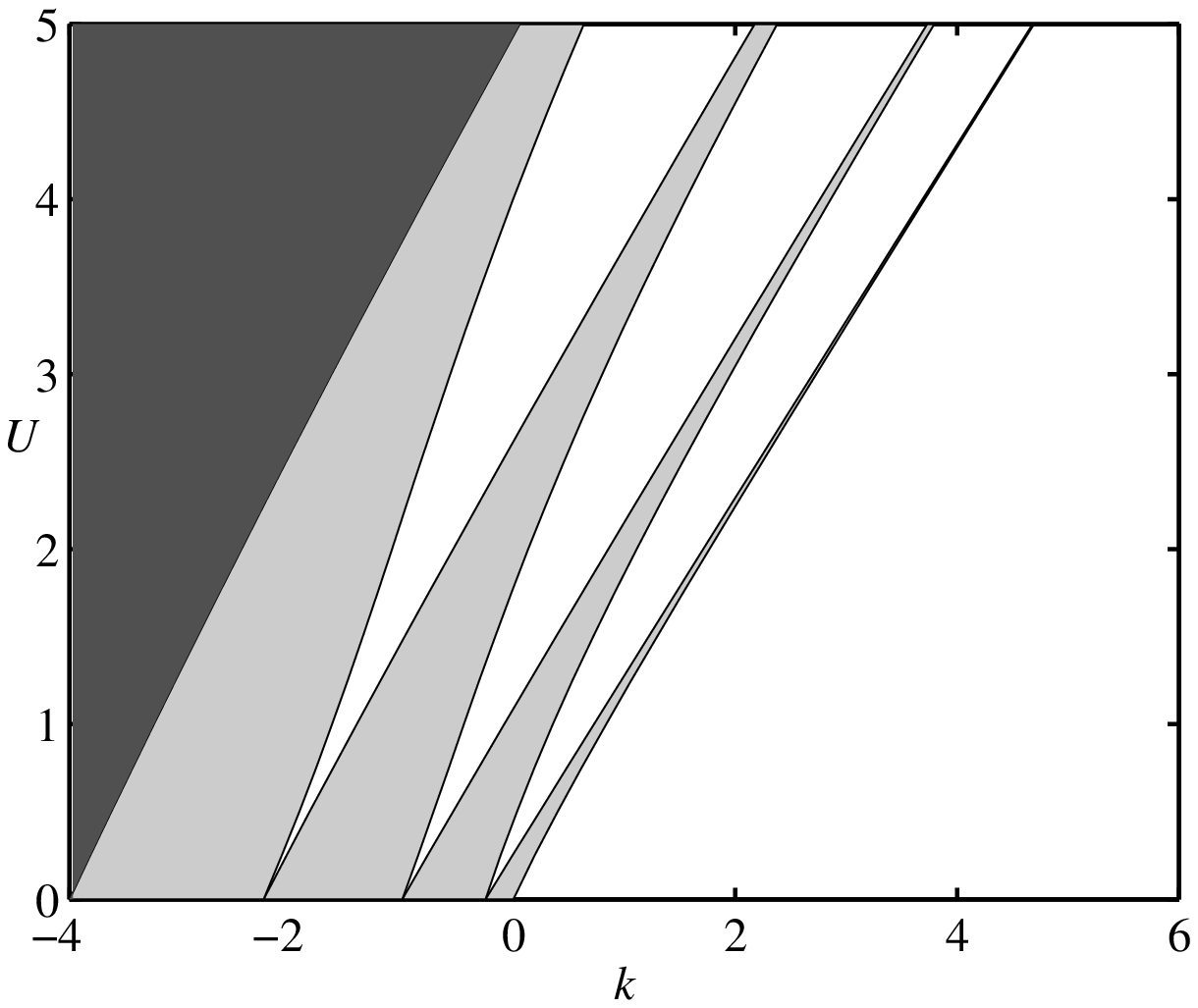}}%
\subfigure[]{\includegraphics[width=3in]{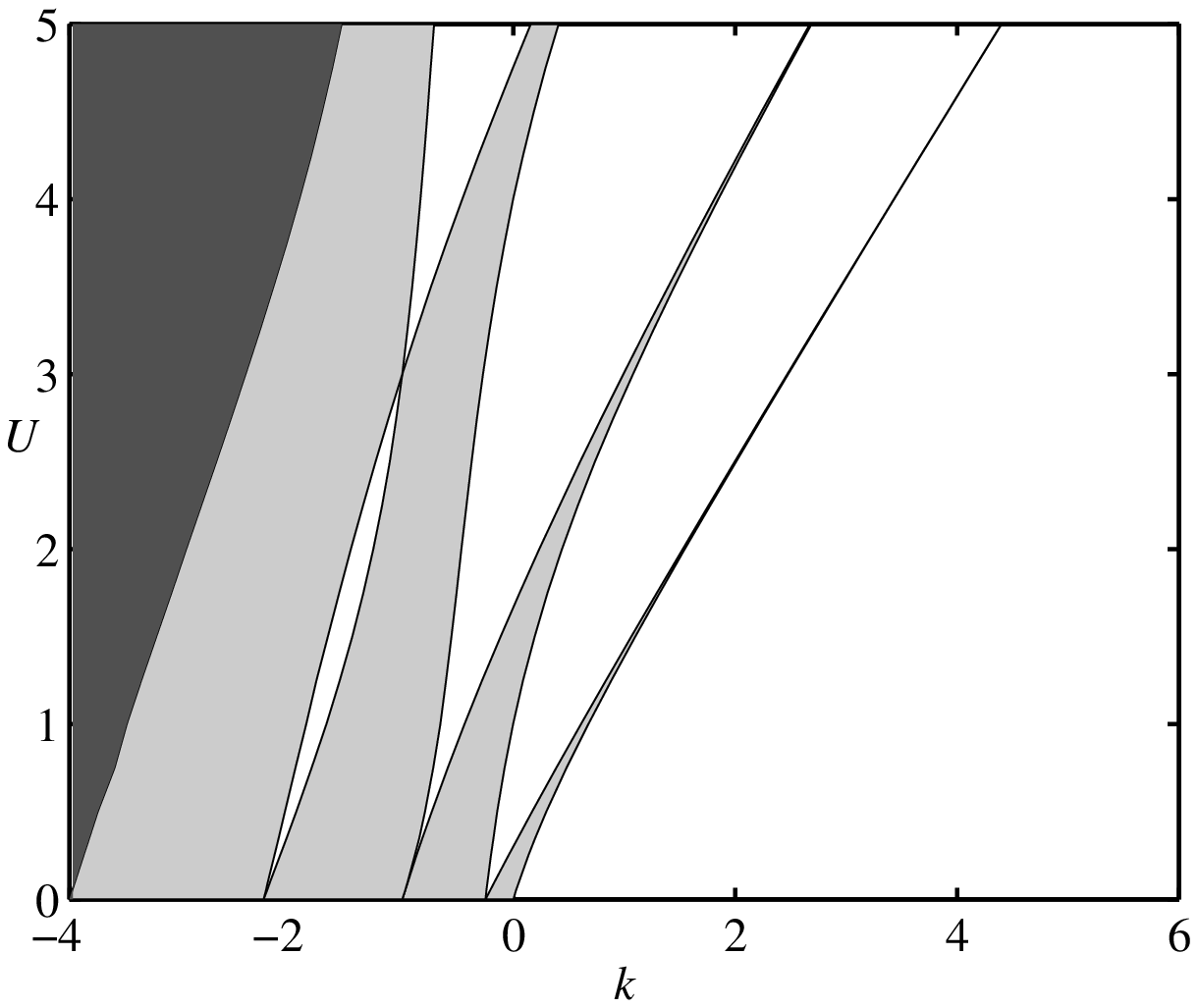}} \subfigure[]{%
\includegraphics[width=3in]{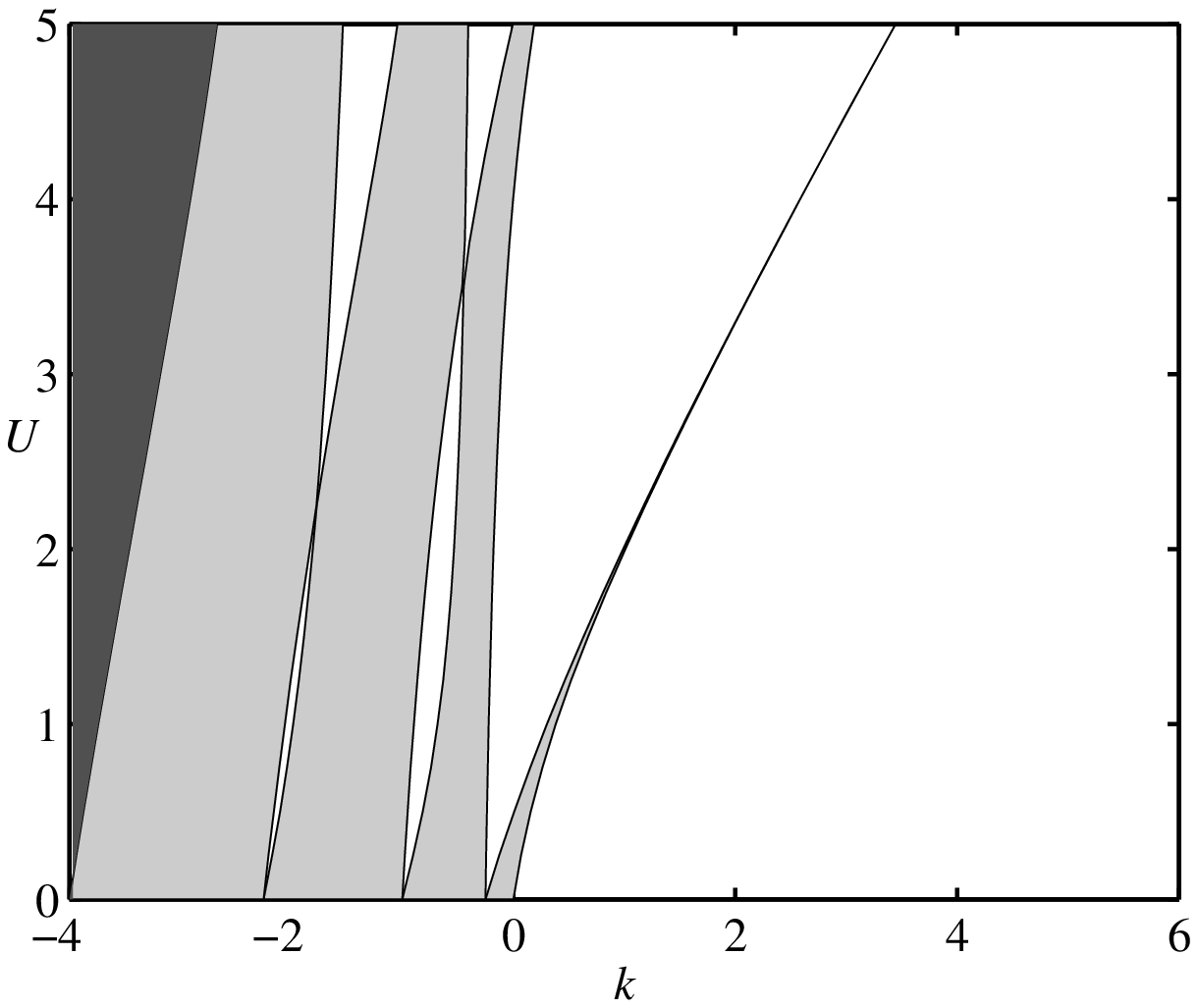}}
\caption{The bandgap structure, as a function of modulation depth $U$ [see
Eq. (\protect\ref{W})], found from the linearization of Eq. (\protect\ref%
{Solution}) for (a) $D=3\protect\pi /2$, (b) $D=\protect\pi $, and (c) $D=%
\protect\pi /2$, the respective ``duty-cycle" values being $\mathrm{DC}%
=0.75,~0.50,~0.25$. Here and in other figures that display ranges of $k$,
shaded areas are occupied by Bloch bands. The spectrum was not computed in
the black area.}
\label{fig1}
\end{figure}

In the SDF model, soliton families were looked for in the first two finite
bandgaps of the spectrum. For very small values of the ``duty cycle",
namely, $\mathrm{DC}\lesssim 0.05$ and sufficiently large values of $U$, the
results are very similar to those reported for the ``Dirac-comb" model in
Refs. \cite{Sukhorukov,Sukhorukov2}, while they are quite different for the
above-mentioned values, $\mathrm{DC}=0.75,~0,50,~0.25$ and $U=3$, for which
systematic results are reported below. Naturally, for fixed $U$ and very
small values of $\mathrm{DC}$, solitons cannot exist with the SDF sign of
the nonlinearity. In particular, fixing $U=3$, we have found that the
solitons disappear at $\mathrm{DC}\leq \left( \mathrm{DC}\right) _{\mathrm{cr%
}}\approx 0.04$. On the other hand, in the SF model the solitons do not
disappear in the same limit, as very narrow solitons with a large total
power can be held in a very narrow channel.

\section{The self-defocusing (SDF) nonlinearity}

\subsection{Fundamental and higher-order spatially symmetric gap solitons}

\begin{figure}[tbph]
\centering\subfigure[]{\includegraphics[width=3in]{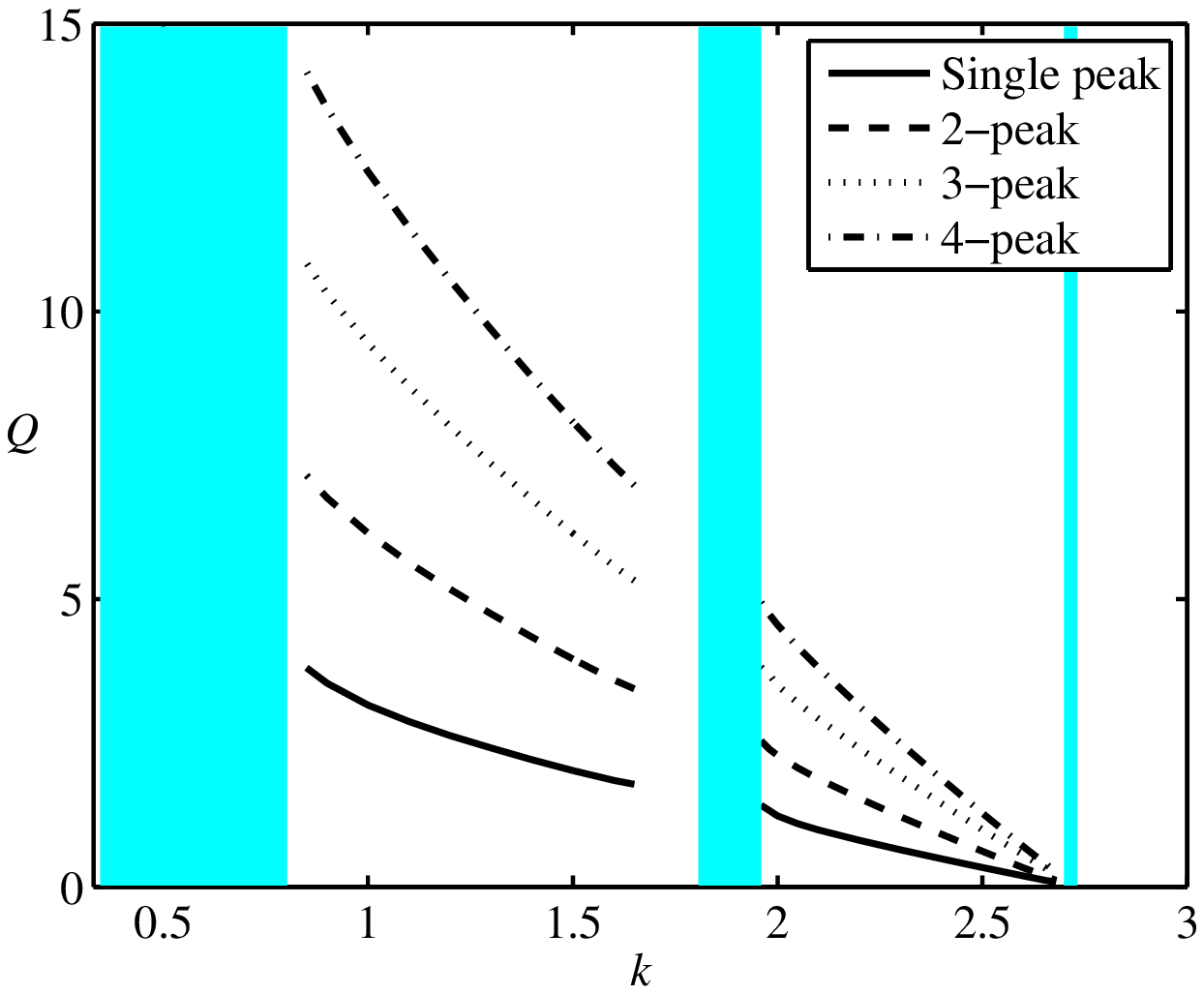}}%
\subfigure[]{\includegraphics[width=3in]{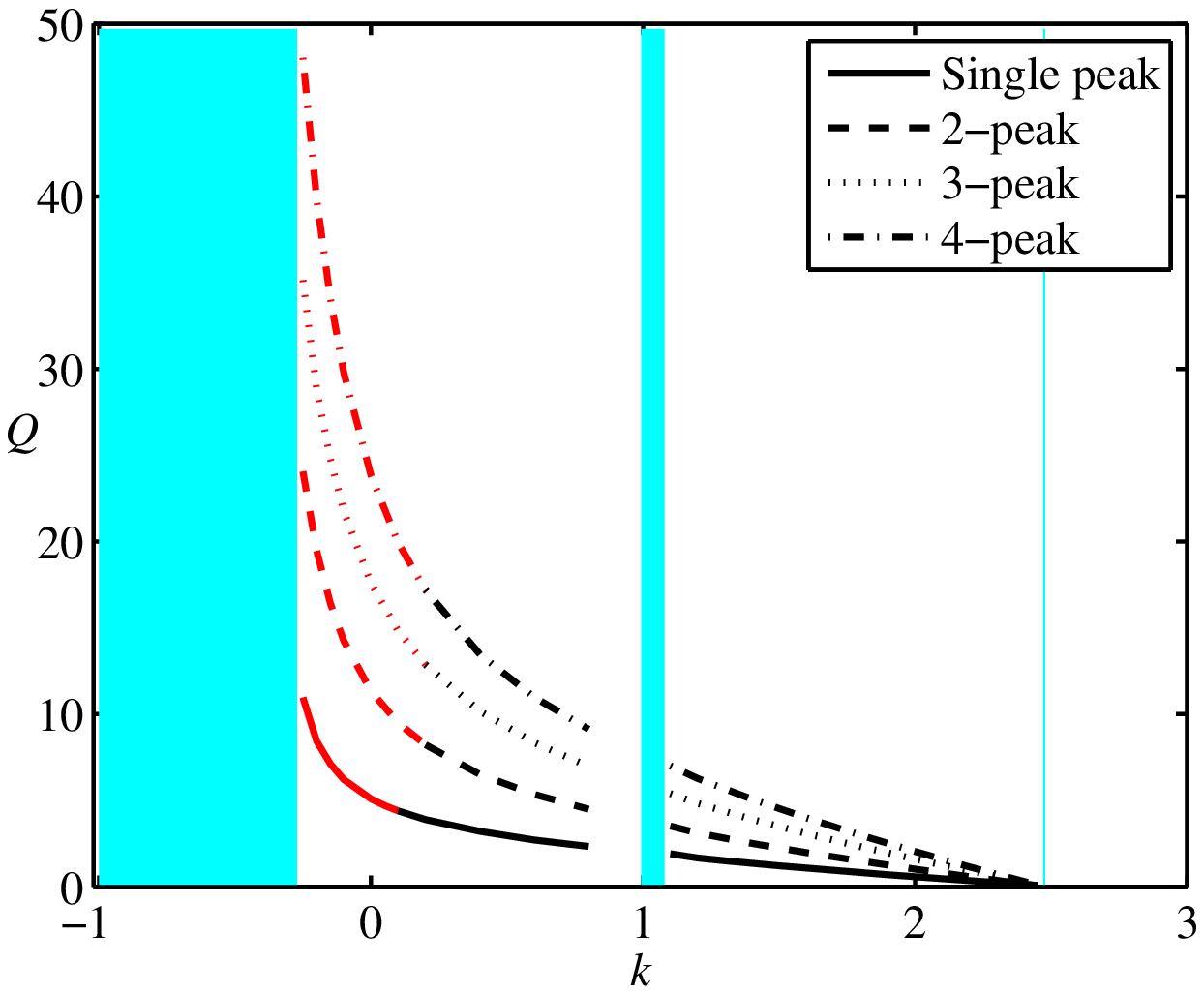}} \subfigure[]{%
\includegraphics[width=3in]{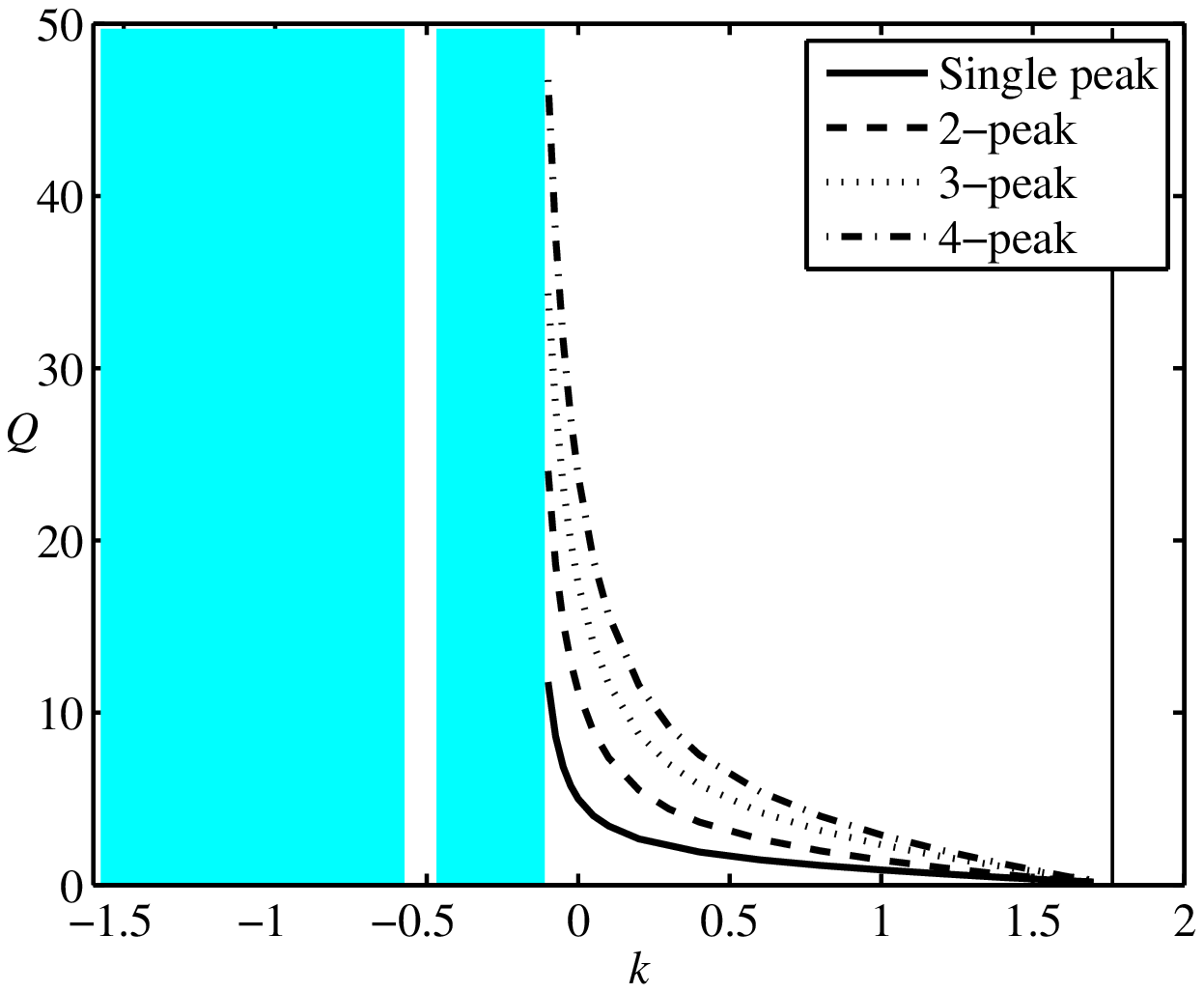}}
\caption{ Total power $Q$ versus propagation constant $k$ for
fundamental and higher-order gap-soliton families in the
defocusing model, is shown at three values of the structural
``duty cycle": $\mathrm{DC}\equiv D/L=0.75$, $0.5$, and $0.25$, in
panels (a), (b), and (c), respectively. Families of stable and
unstable solutions are shown, severally, by black and red (gray,
in the black-and-white version) curves.} \label{fig2}
\end{figure}

Different families of GSs are identified by the number of peaks in them. In
addition to the fundamental (single-peak) solitons, we have found families
of spatially symmetric (even) localized modes with two, three, and four
peaks. Figure \ref{fig2} represents these families by means of respective
dependences $Q(k)$, at the three above-mentioned characteristic values of $%
\mathrm{DC}$, in the first two finite bandgaps. In the case of $\mathrm{DC}%
=0.25$, the second bandgap is empty, as no soliton solutions could be found
in it. It is also necessary to mention that, in the latter case, the
higher-order solitons are classified as $2$-, $3$-, and $4$-peak ones by
analogy with the situation observed for $\mathrm{DC}=0.75$ and $0.50$, while
their actual shape is different, see Fig. \ref{fig9} below (in fact, numbers
$2$, $3$ and $4$ for these solitons pertain not to the number of peaks but
rather to the number of channels occupied by each localized mode).

For $\mathrm{DC}=0.75$, the GS families are entirely stable, while, in the
intermediate case, $\mathrm{DC}=0.50$, the analysis identifies an intrinsic
stability border in the middle of the second bandgap. The transition from
stable to unstable solutions occurs at points where the GSs feature a
specific \textit{flat-top} shape, see below.

\begin{figure}[tbp]
\centering\subfigure[]{\includegraphics[width=4in]{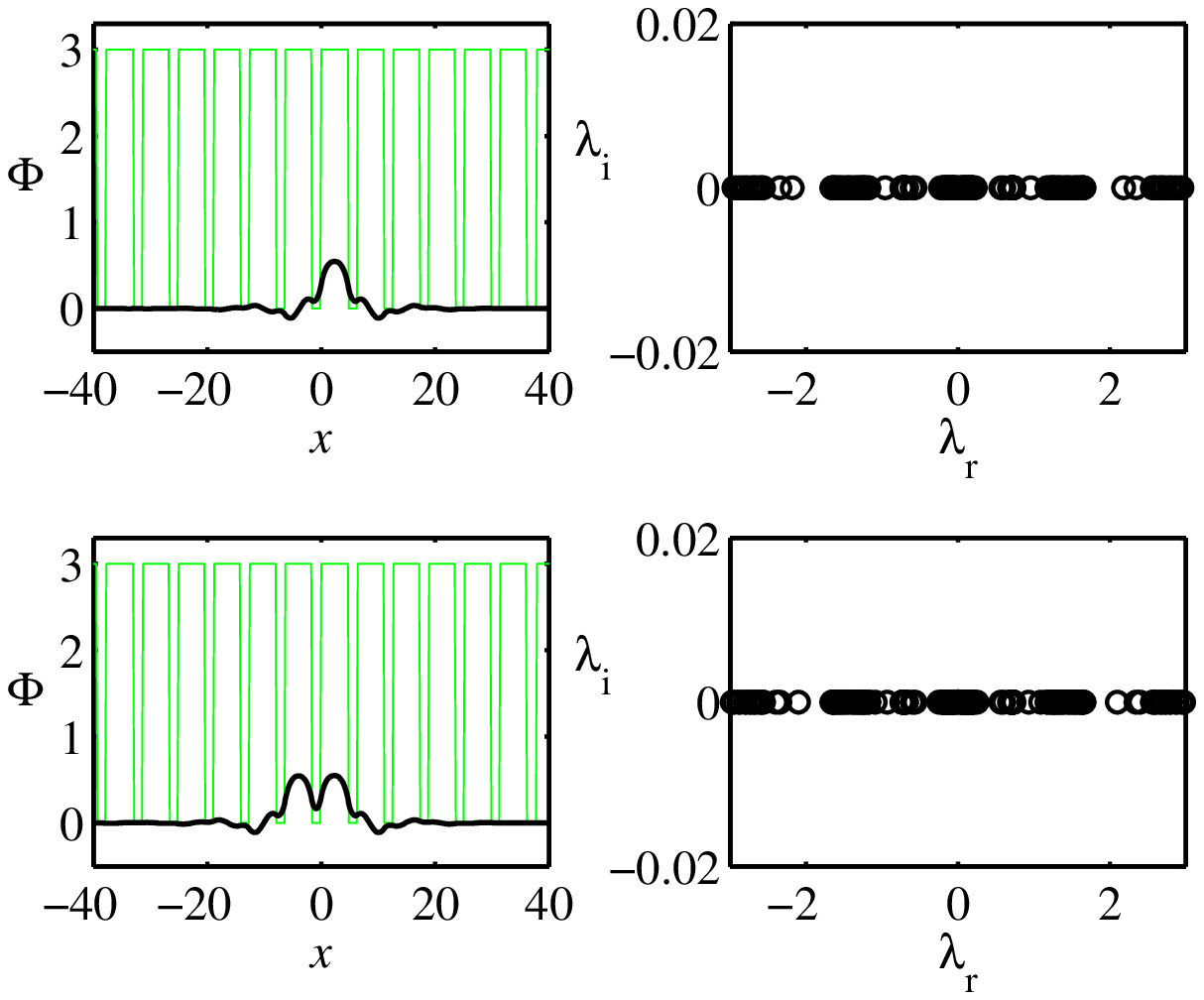}} %
\subfigure[]{\includegraphics[width=4in]{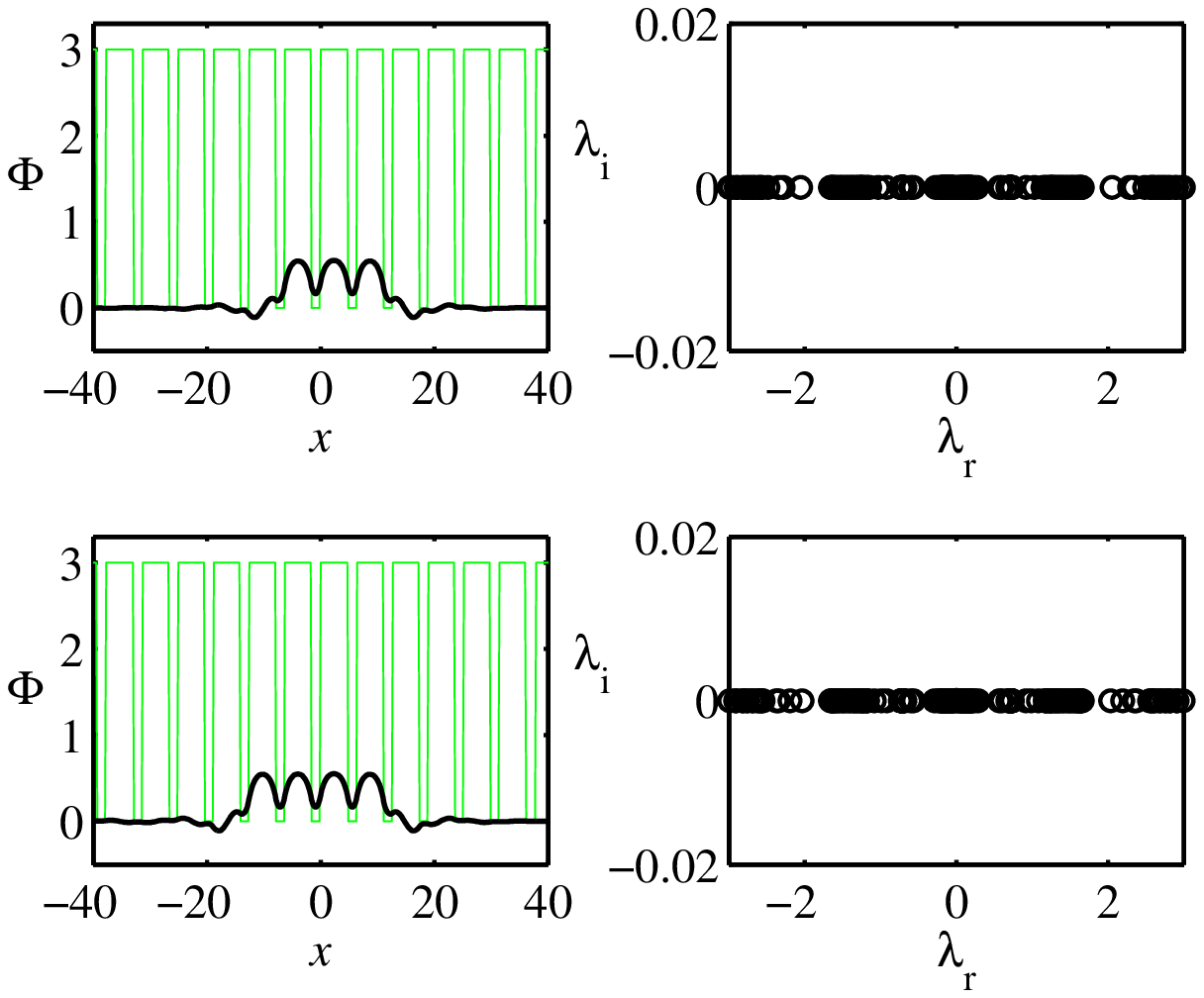}} \caption{
Examples of stable fundamental and double-peak solitons (a), and
their three- and four-peak counterparts (b), found in the first
bandgap of the defocusing model with $\mathrm{DC}=0.75$, for a
fixed value of the propagation constant, $k=2$. In each panel
(here and in similar figures below), the right plot shows the
spectral plane of stability
eigenvalues for the respective soliton, $\protect\lambda \equiv \protect%
\lambda _{\mathrm{r}}+i\protect\lambda _{\mathrm{i}}$ (the stability in
implied by $\protect\lambda _{\mathrm{i}}=0$). Here and in other figures,
the background pattern represents the underlying structure of the layered
medium. }
\label{fig3}
\end{figure}

\begin{figure}[tbp]
\centering\subfigure[]{\includegraphics[width=4in]{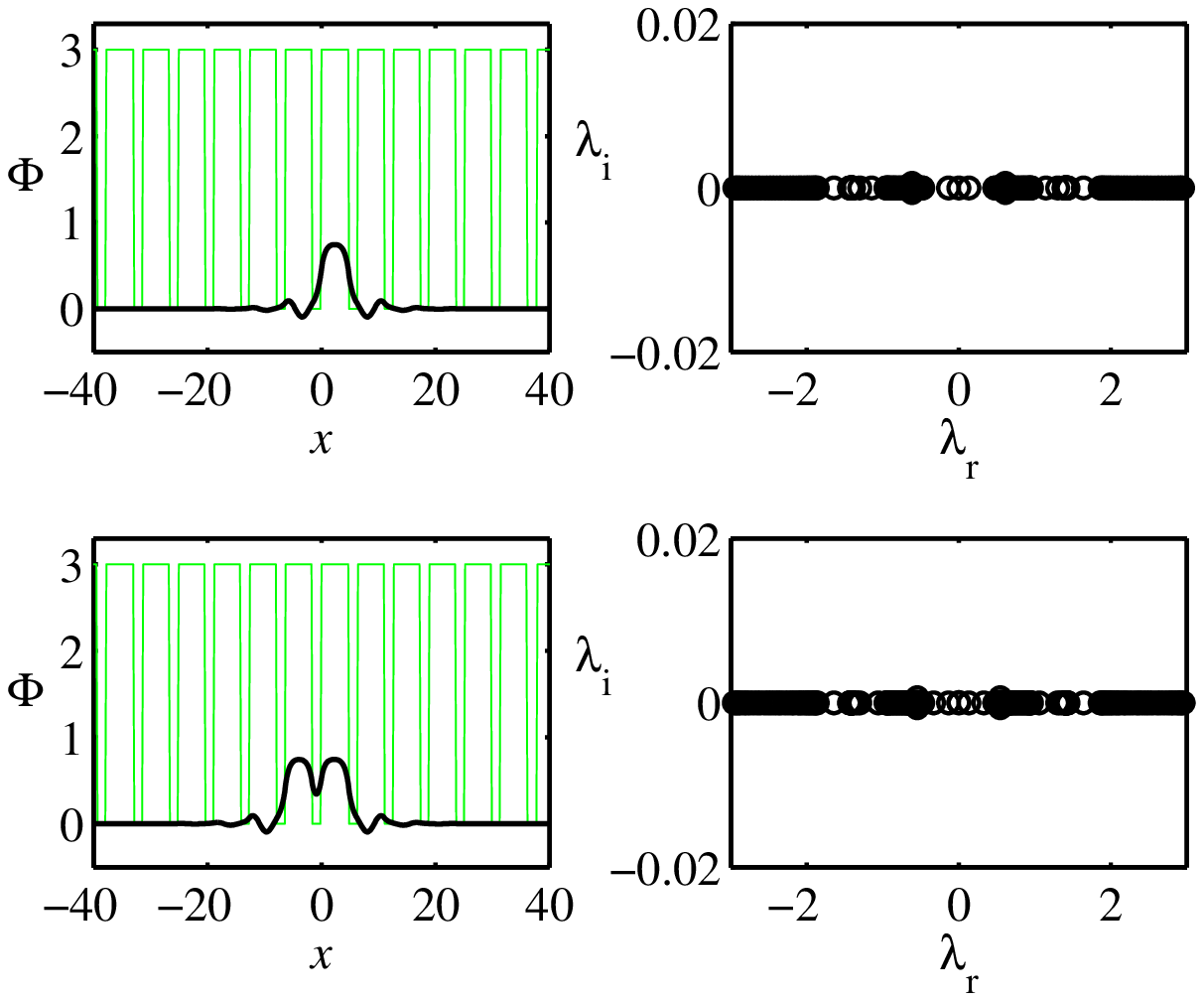}} %
\subfigure[]{\includegraphics[width=4in]{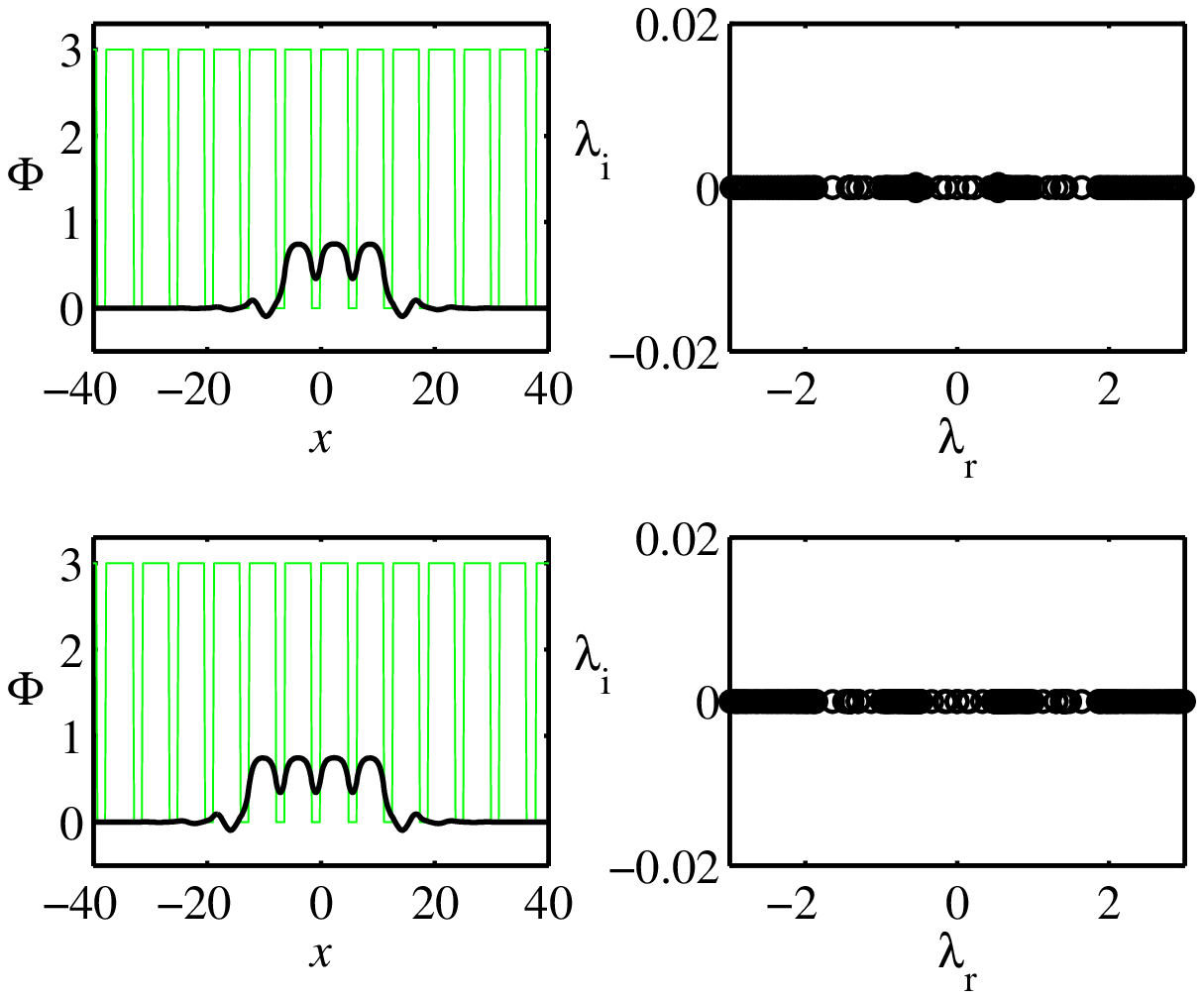}} \caption{ The
same as in Fig. 4, but in the second bandgap, for $k=1.3$.}
\label{fig4}
\end{figure}

For $\mathrm{DC}=0.75$, typical examples of stable fundamental GSs, together
with $2$-, $3$- and $4$-peak solitons, are displayed in Fig. \ref{fig3} and %
\ref{fig4}, in the first and second bandgaps, respectively. For both $%
\mathrm{DC}=0.75$ and $\mathrm{DC}=0.50$, maxima and minima of the local
power in \emph{stable} solitons are always observed, respectively, in
waveguiding channels and in empty layers between them, while for $\mathrm{DC}%
=0.25$ the situation is different, see below. Another relevant conclusion is
that the $N$-peak solitons may be clearly considered as bound states of $N$
fundamental ones, as suggested by the fact that, as obvious in Fig. 2, the
respective powers are related by $Q_{N}\approx NQ_{\mathrm{fund}}$.

\begin{figure}[tbp]
\centering\subfigure[]{\includegraphics[width=4in]{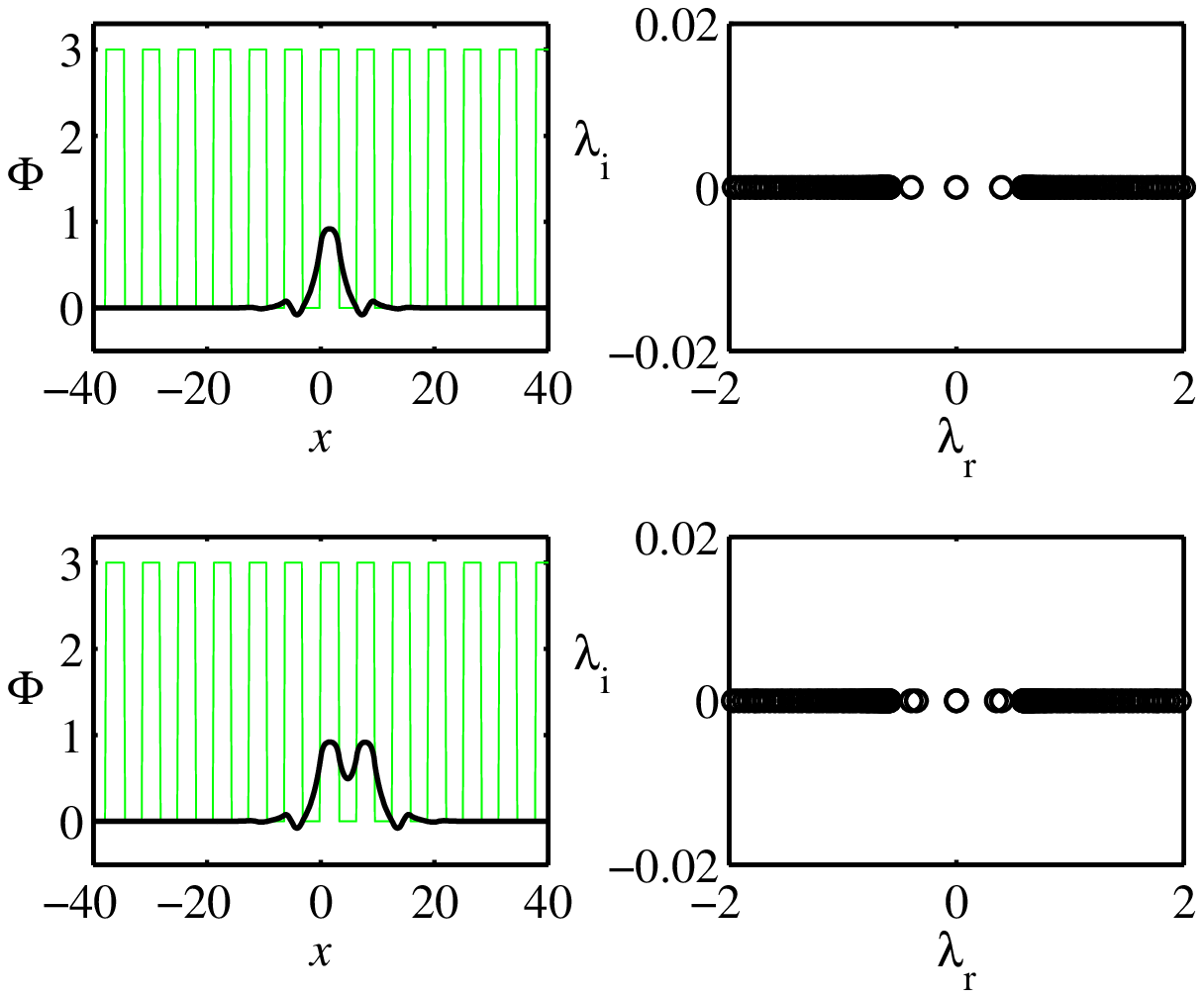}} %
\subfigure[]{\includegraphics[width=4in]{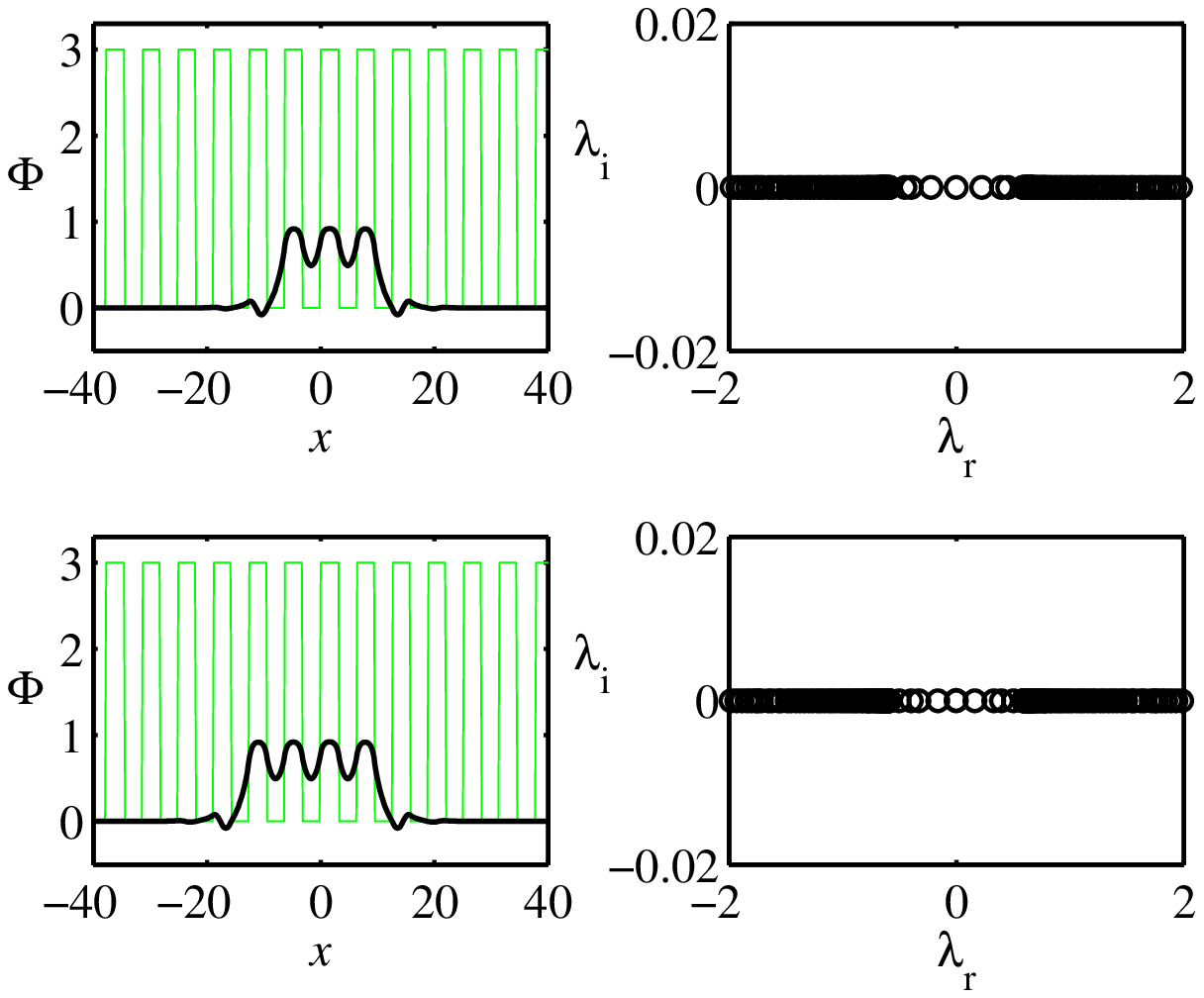}} \caption{
Examples of stable solitons in the second bandgap of the
defocusing model with $\mathrm{DC}=0.5$, for a fixed propagation
constant, $k=0.4$: (a) a fundamental soliton, with $Q=3.20$, and a
double-peak one, with $Q=6.69$; (b) a three- and four-peak solitons, with $%
Q=10.17$ and $Q=13.65$, respectively.}
\label{fig5}
\end{figure}

In fact, dependences $Q(k)$ and shapes of the solitons in the SDF model with
$\mathrm{DC}=0.75$ are very similar to those known for ordinary GSs in the
standard model with the KP potential and uniform repulsive nonlinearity, cf.
Refs. \cite{Smerzi,Carr,Gisin}. For $\mathrm{DC}=0.5$, the fundamental and
multi-peak GSs in the first bandgap are close to their counterparts
displayed in Fig. \ref{fig3} for $\mathrm{DC}=0.75$. However, as shown in
Fig. \ref{fig5}, they are essentially different in the second bandgap of the
model with $\mathrm{DC}=0.5$ (hence, different from the ordinary GSs too),
showing a trend to flattening.

\begin{figure}[tbp]
\centering\subfigure[]{\includegraphics[width=4in]{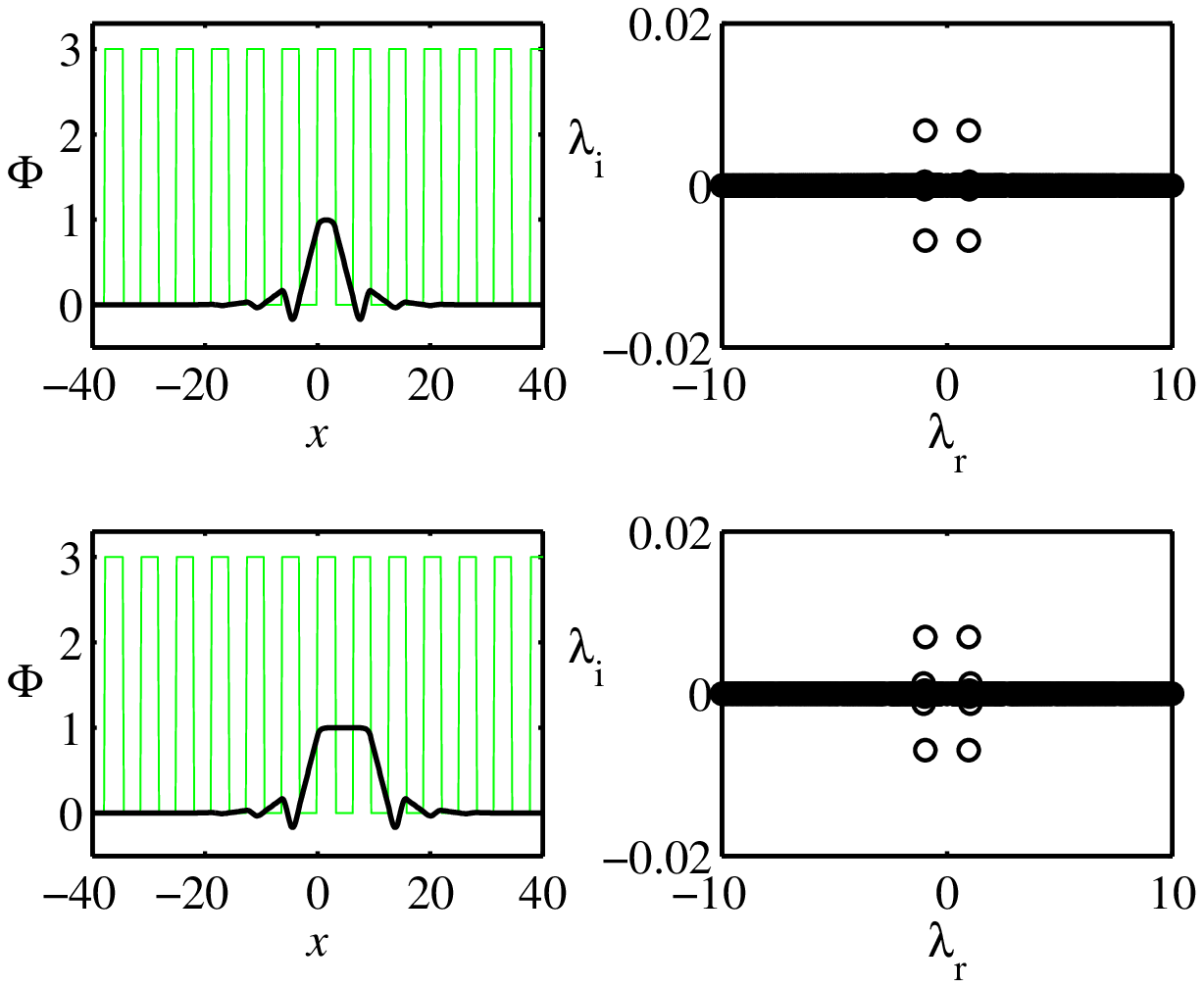}} %
\subfigure[]{\includegraphics[width=4in]{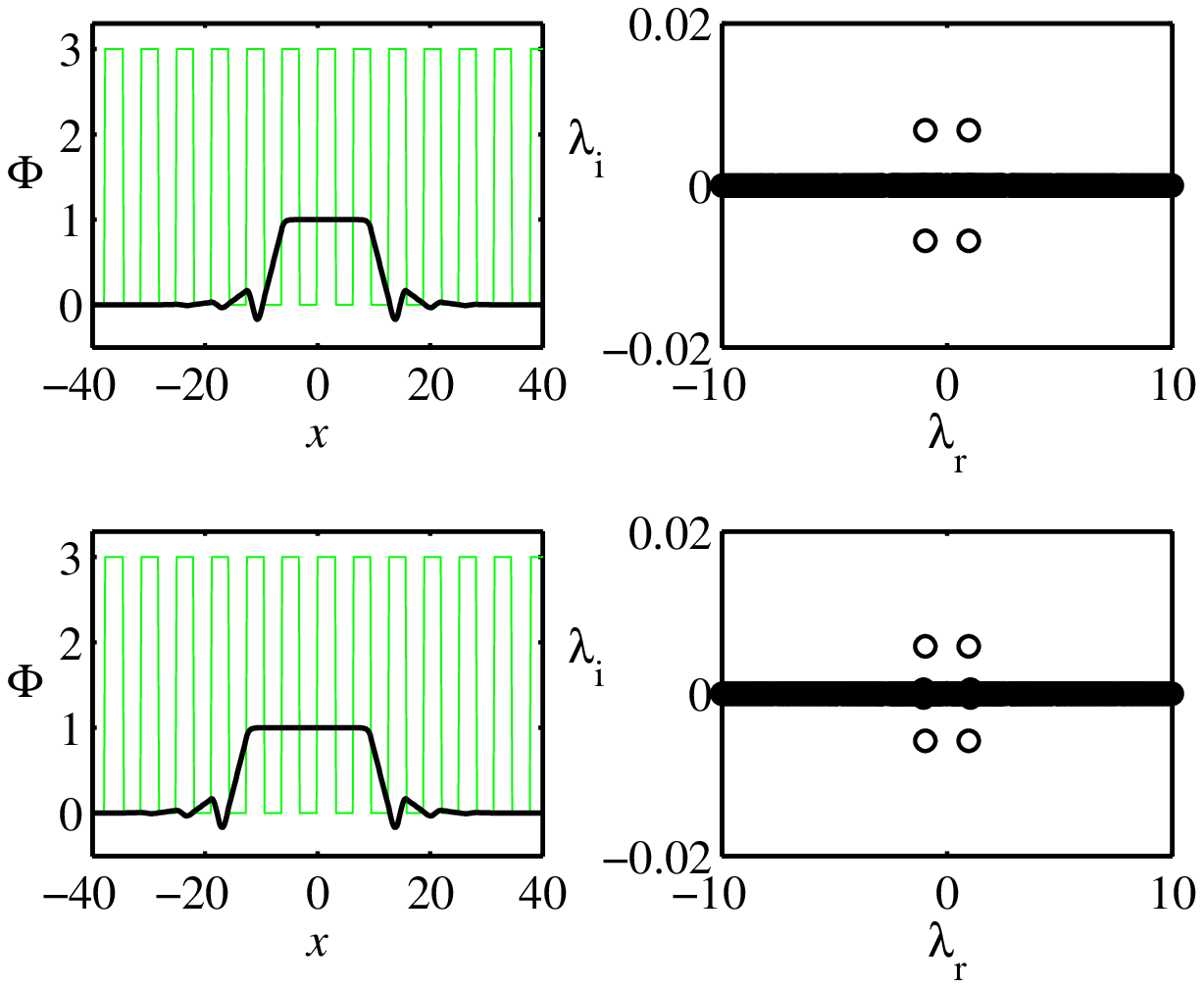}}
\caption{ Weakly unstable nearly flat-top solitons found at $%
k=0$ in the second bandgap of the defocusing model with\textrm{\ }$\mathrm{CD%
}=0.50$: (a) A fundamental soliton, with $Q=5.08$, and flat-top counterpart
of the double-peak soliton, with $Q=11.33$. (b) Flat-top counterparts of
three- and four-peak solitons, with $Q=17.58$ and $23.83$, respectively.}
\label{fig6}
\end{figure}

As said above, a new feature found in the second bandgap of the SDF model
with $\mathrm{DC}=0.5$ is destabilization of all branches of the GSs
(fundamental and multi-peak ones alike). As seen in Fig. 2(b), this happens
at small positive values of $k$. Exactly at the stability border (and close
to it, on both sides) all the solitons feature a flat-top shape, as shown in
Fig. \ref{fig6}.

\begin{figure}[tbp]
\centering\subfigure[]{\includegraphics[width=4in]{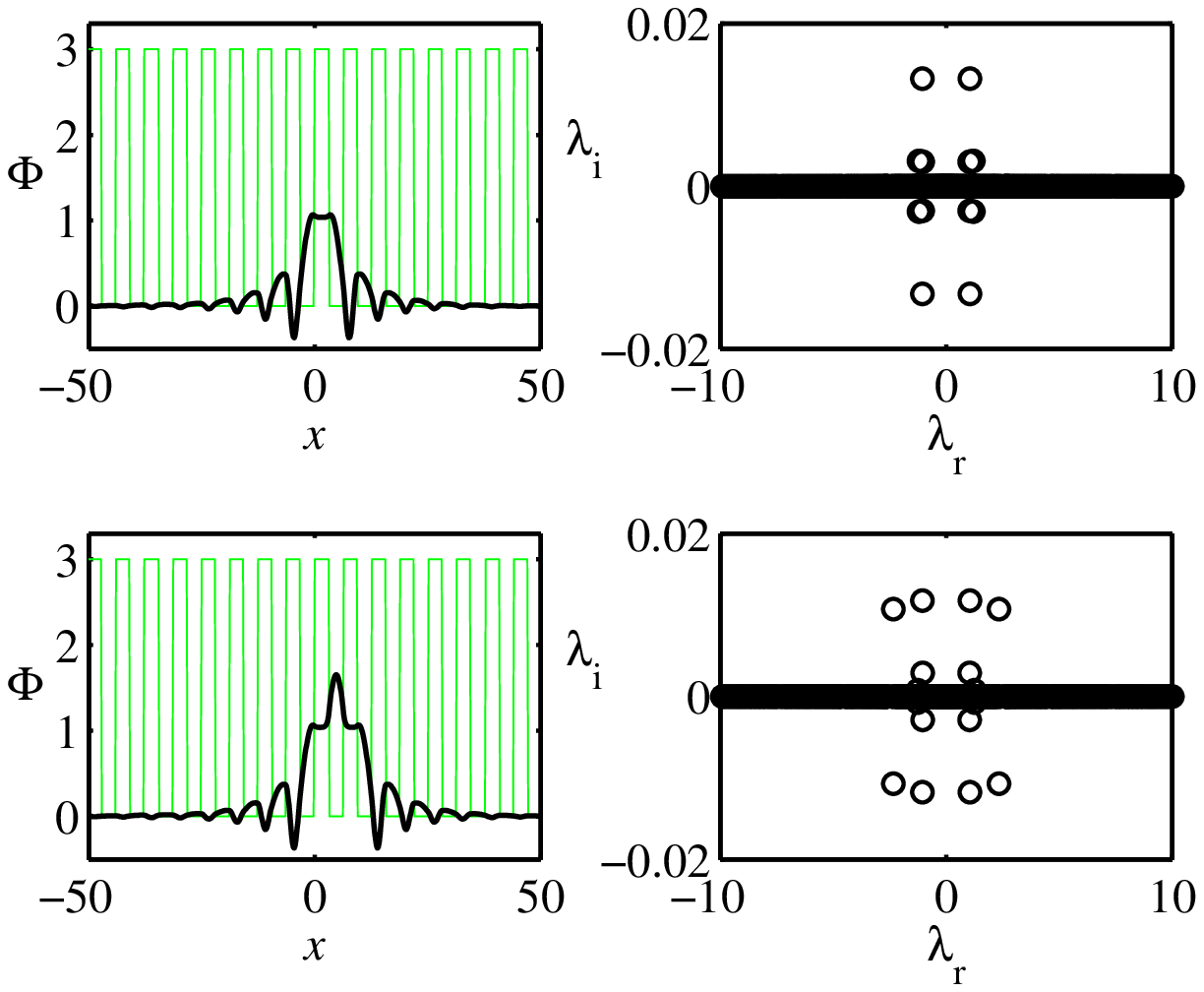}} %
\subfigure[]{\includegraphics[width=4in]{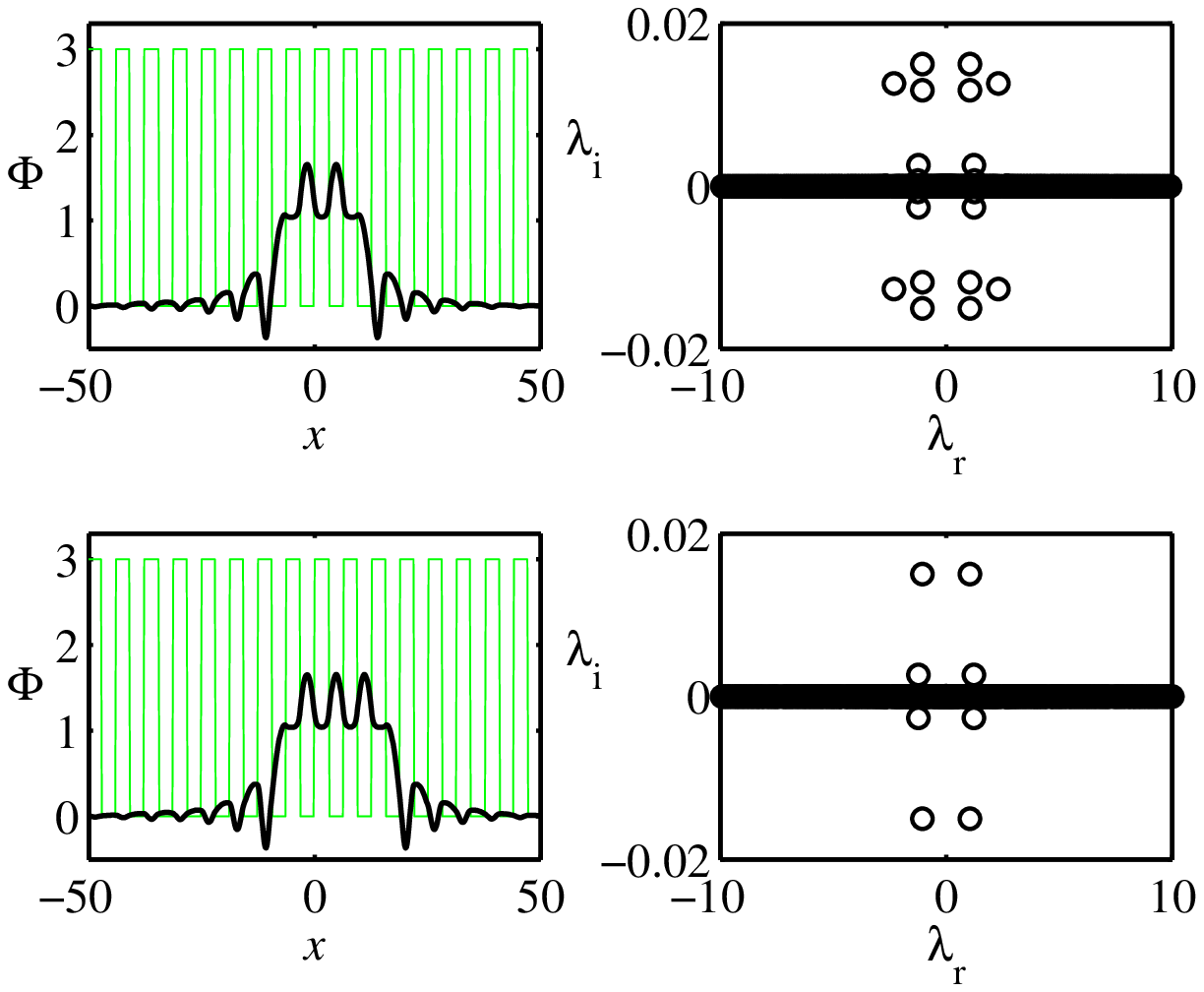}} \caption{
Unstable solitons with the \textit{inverted shape} (i.e., with
local power maxima in empty layers). All the examples are shown
for $k=-0.22$. The top and bottom plots in panel (a) display,
respectively, a fundamental soliton, with $Q=9.17$ , and the
former double-peak soliton (which actually features a single-peak
structure, as a result of the inversion), with $Q=21.06$. The top
and bottom plots in panel (b) display, respectively, inverted
counterparts of former three- and four-peak solitons, with powers
$Q=30.88$ and $Q=42.74$.} \label{fig7}
\end{figure}

Deeper into the instability region, the higher-order GSs develop \textit{%
inverted} \textit{shapes}, with respect to their stable counterparts, see
Fig. \ref{fig7}: on top of the flat-top background, peaks emerge in empty
spaces between the waveguiding channels. In this case, the fundamental
soliton tends to keep a nearly flat-top shape (although its instability is
not essentially weaker than that of its higher-order counterparts),
featuring a shallow depression in the central guiding channel. Thus, each
localized mode that had $N$ peaks, and $N-1$ troughs between them, in the
stable configuration, develops $N-1$ inverted peaks, in locations of the
former troughs, after passing the destabilization threshold (which
corresponds to the flat-top configuration). In fact, this shape
transformation suggests that, with the increase of total power $Q$, the
effective repulsive nonlinear potential becomes stronger than the trapping
linear one, pushing the wave into the empty space and destabilizing the
trapped state.

\begin{figure}[tbp]
\centering\subfigure[]{\includegraphics[width=4in]{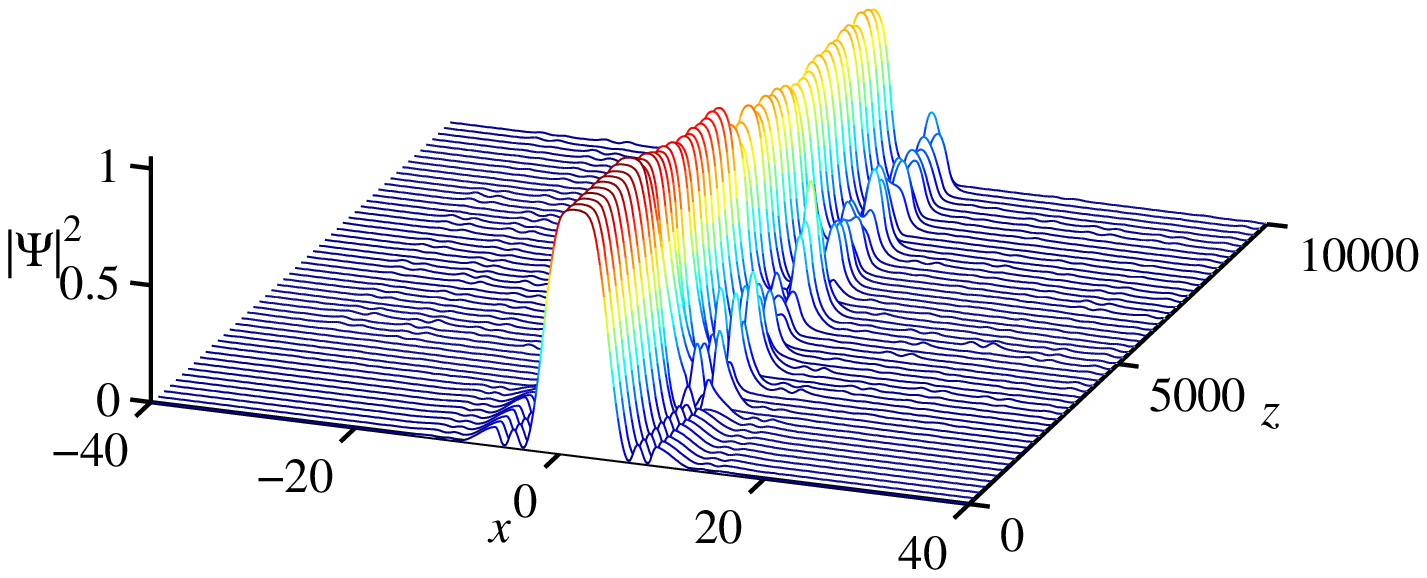}} %
\subfigure[]{\includegraphics[width=4in]{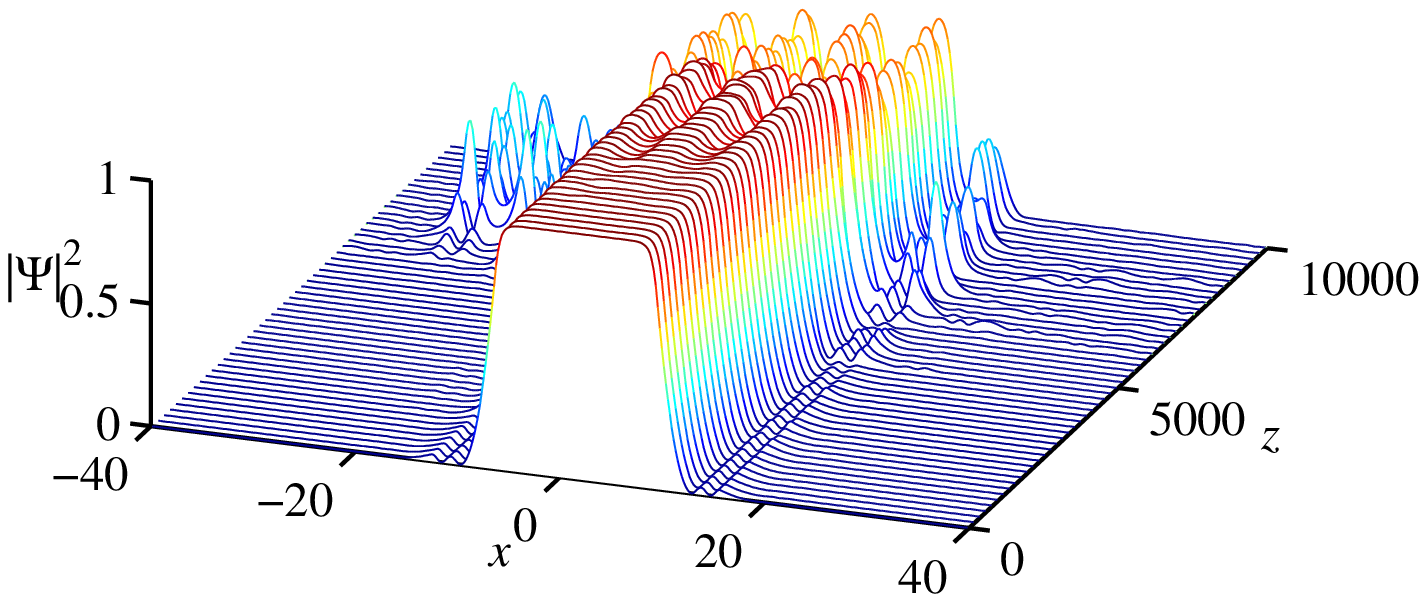}} \caption{ (a)
The evolution of an unstable fundamental soliton
with $k=-0.15$ and $Q=17.58$ in the self-defocusing model with $\mathrm{DC}%
=0.5$. (b) The same for an unstable flat-top soliton belonging to the
three-peak family, with $k=0$ and $Q=17.58$.}
\label{fig8}
\end{figure}

\begin{figure}[tbph]
\centering\subfigure[]{\includegraphics[width=4in]{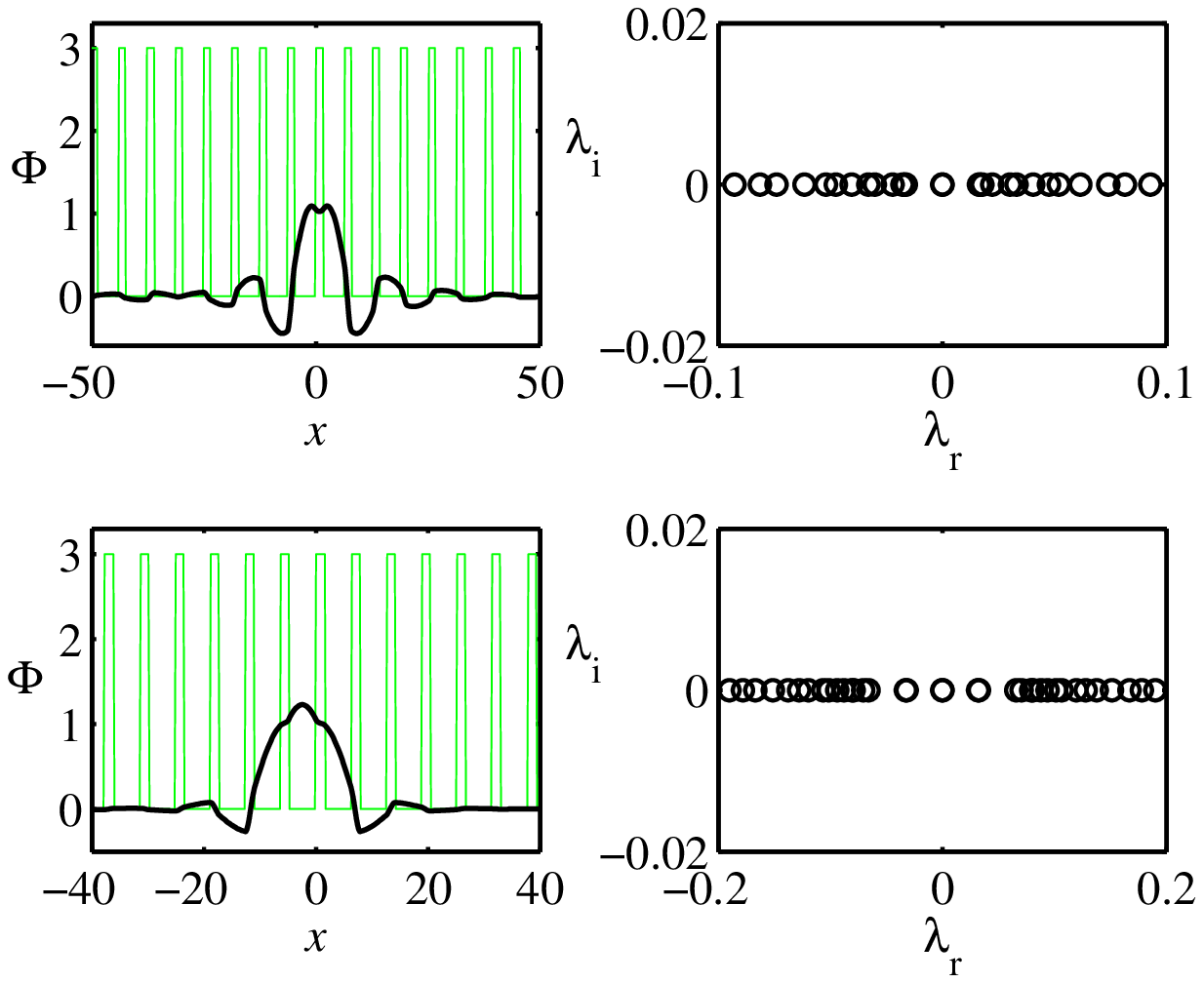}} %
\subfigure[]{\includegraphics[width=4in]{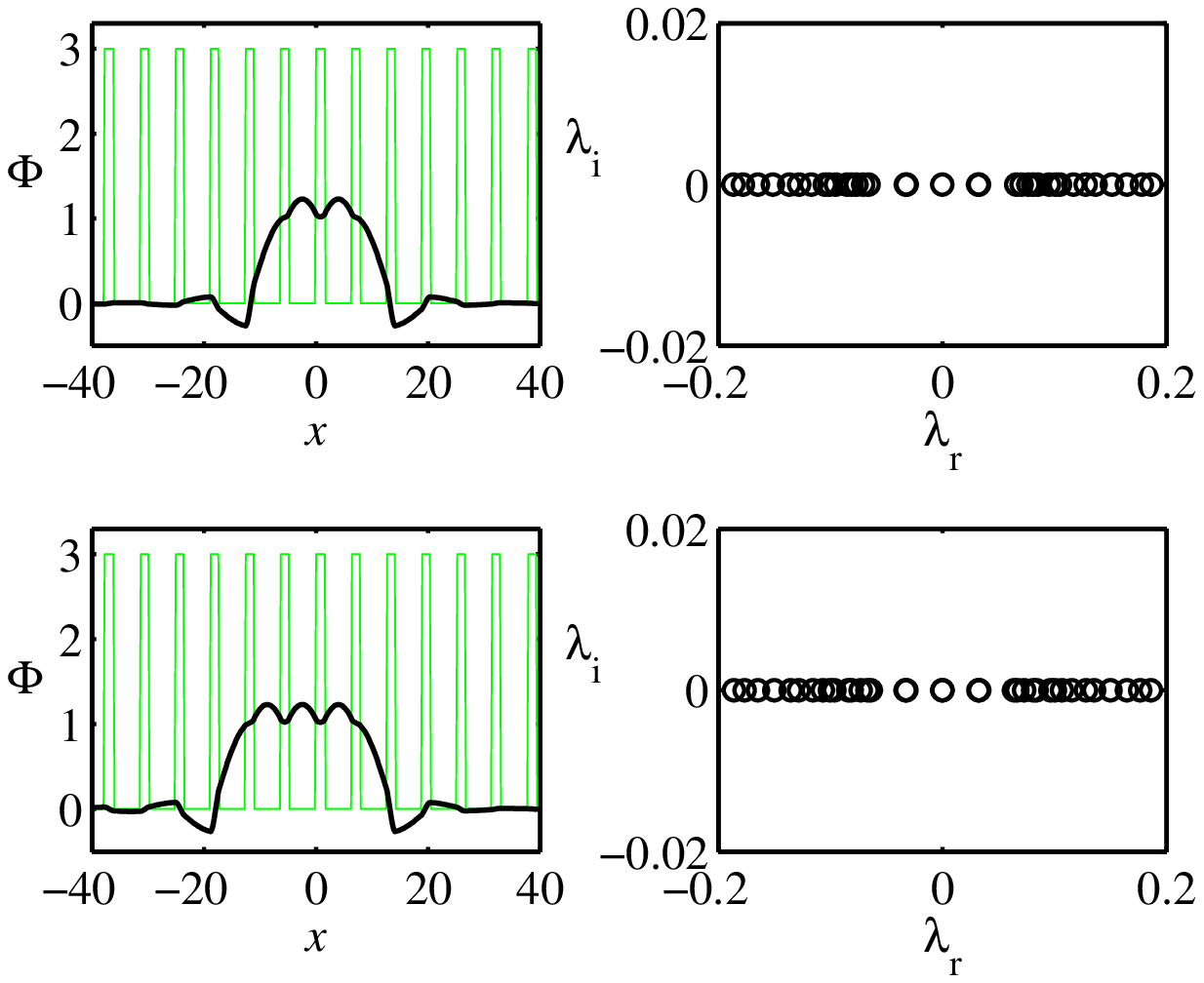}} \caption{
Examples of stable gap solitons in the first bandgap of the
self-defocusing model with $\mathrm{DC}=0.25$. In panel (a), the
top and bottom plots represent, respectively, a fundamental
soliton (i.e., the one localized in a single channel), with
$k=-0.1$ and $Q=11.77$, and a soliton occupying two channels, with
$k=-0.05$ and $Q=15.26$. Panel (b)
displays higher-order solitons occupying three and four channels, with $%
Q=23.18$ and $31.60$, respectively, both pertaining to $k=-0.05$. Note that
these stable solitons feature local maxima of the power in empty channels.}
\label{fig9}
\end{figure}

When the flat-top fundamental GS becomes unstable, direct simulations
demonstrate that growing perturbations slowly destroy it, see Fig. 8(a). As
for unstable flat-top counterparts of stable multi-peak solitons, the
perturbation first splits them back into a multi-peak structure, and then
gradually destroys them. A typical example of this scenario of the
instability development is shown, for the GS of the $3$-peak type, in Fig. %
\ref{fig8}(b).

At small values of $\mathrm{DC}$, such as $\mathrm{DC}=0.25$, GSs of all
types are found only in the first bandgap. A peculiarity of the soliton
families in this case is that they feature peaks in empty layers, being,
nevertheless, \emph{completely stable}, on the contrary to their unstable
counterparts in the second bandgap found at $\mathrm{DC}=0.50$, which
demonstrate a similar structure, as described above. Examples of these
stable solitons are displayed in Fig. \ref{fig9}.

\subsection{Antisymmetric gap solitons}

\begin{figure}[h]
\centering\subfigure[]{\includegraphics[width=3in]{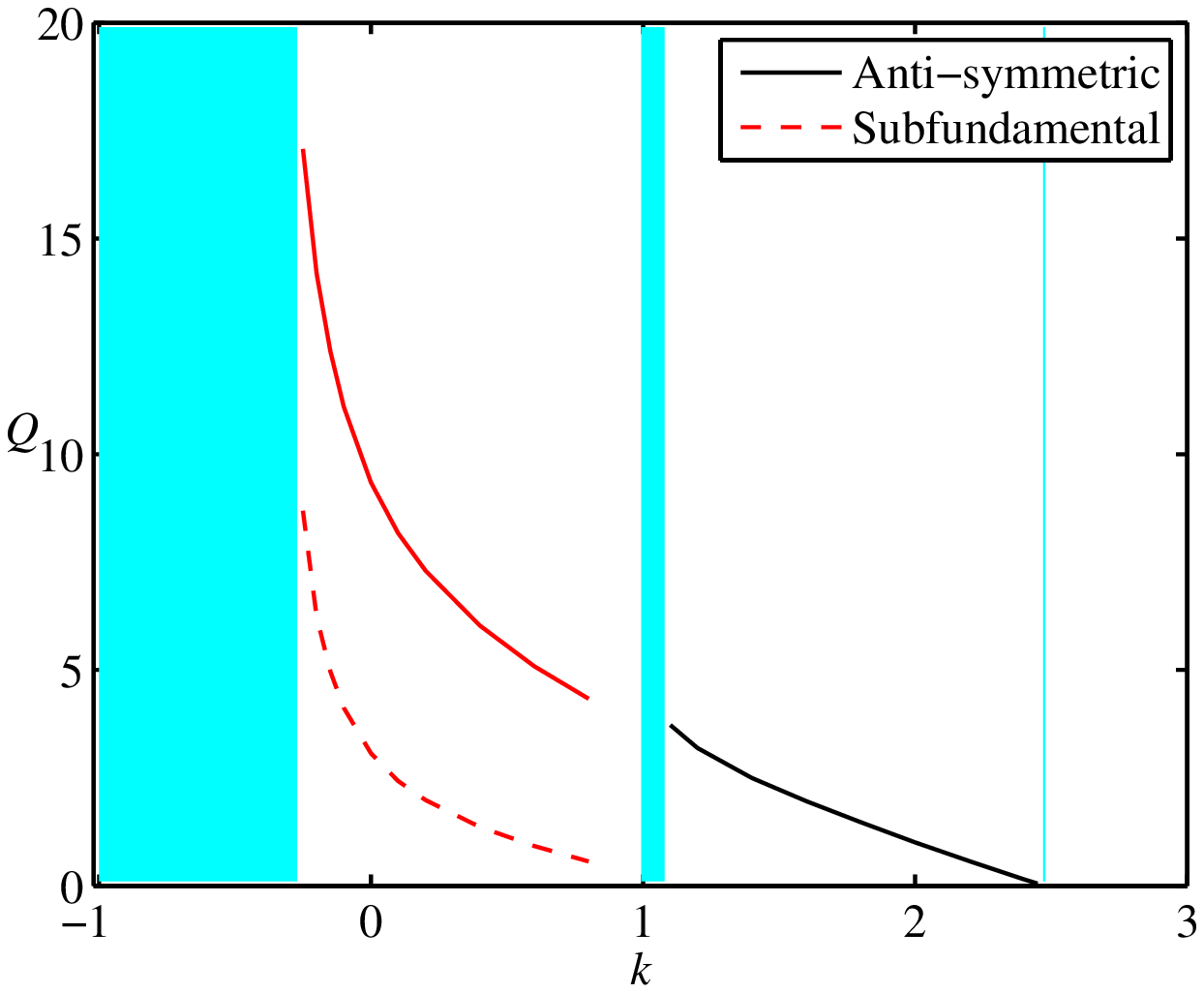}} %
\subfigure[]{\includegraphics[width=4in]{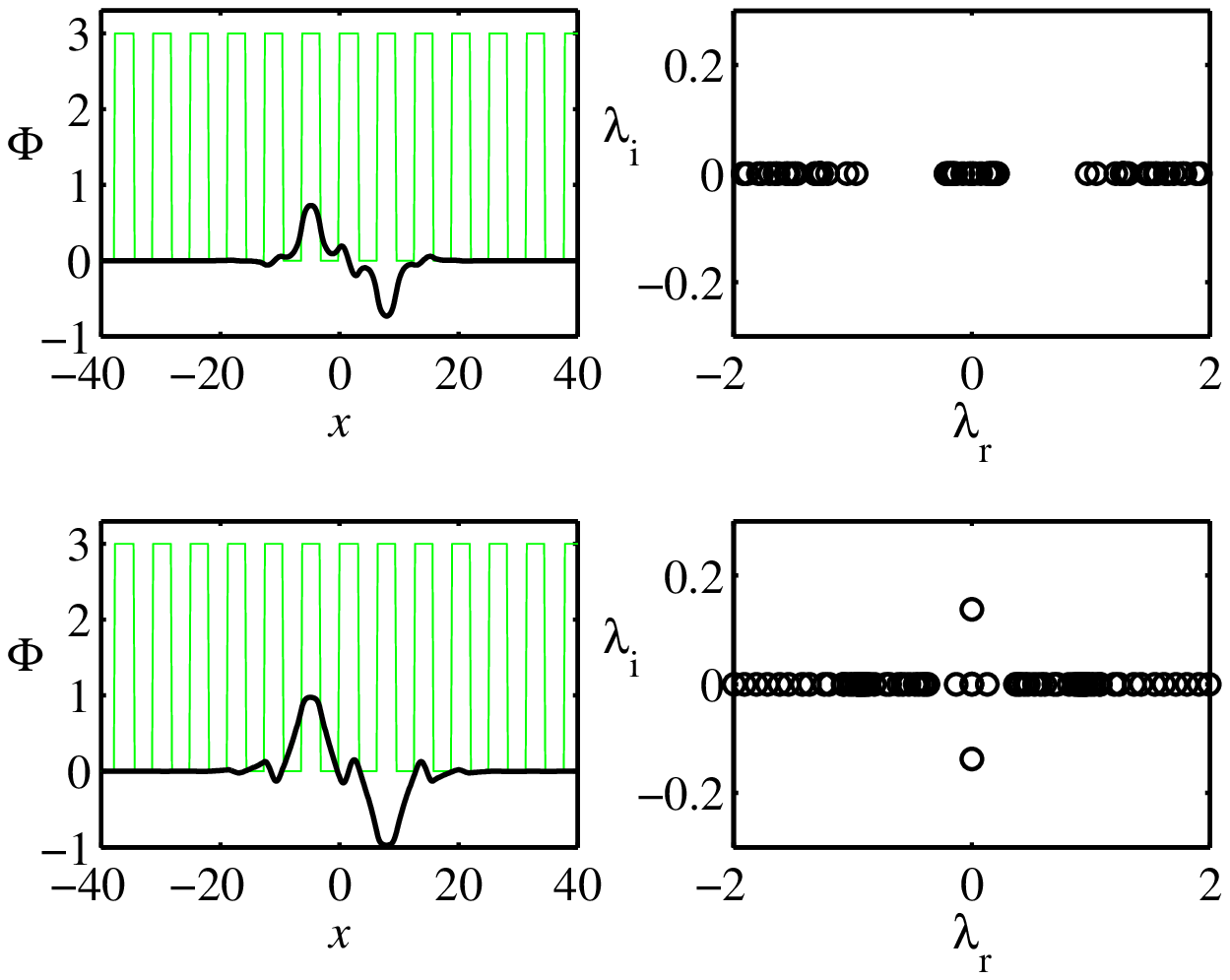}} \caption{
(a) The total power of antisymmetric (\textquotedblleft twisted")
bound states of fundamental gap solitons versus the propagation
constant, in the defocusing model with $\mathrm{DC}=0.50$. The
same characteristic is also shown for the family of subfundamental
solitons in the second bandgap. Typical examples of stable ($k=1.2$ and $%
Q=3.19$) and unstable ($k=0.1$ and $Q=8.17$) antisymmetric bound states, in
the first and second bandgaps, are shown in panel (b). }
\label{fig10}
\end{figure}

\begin{figure}[h]
\centering\subfigure[]{\includegraphics[width=4in]{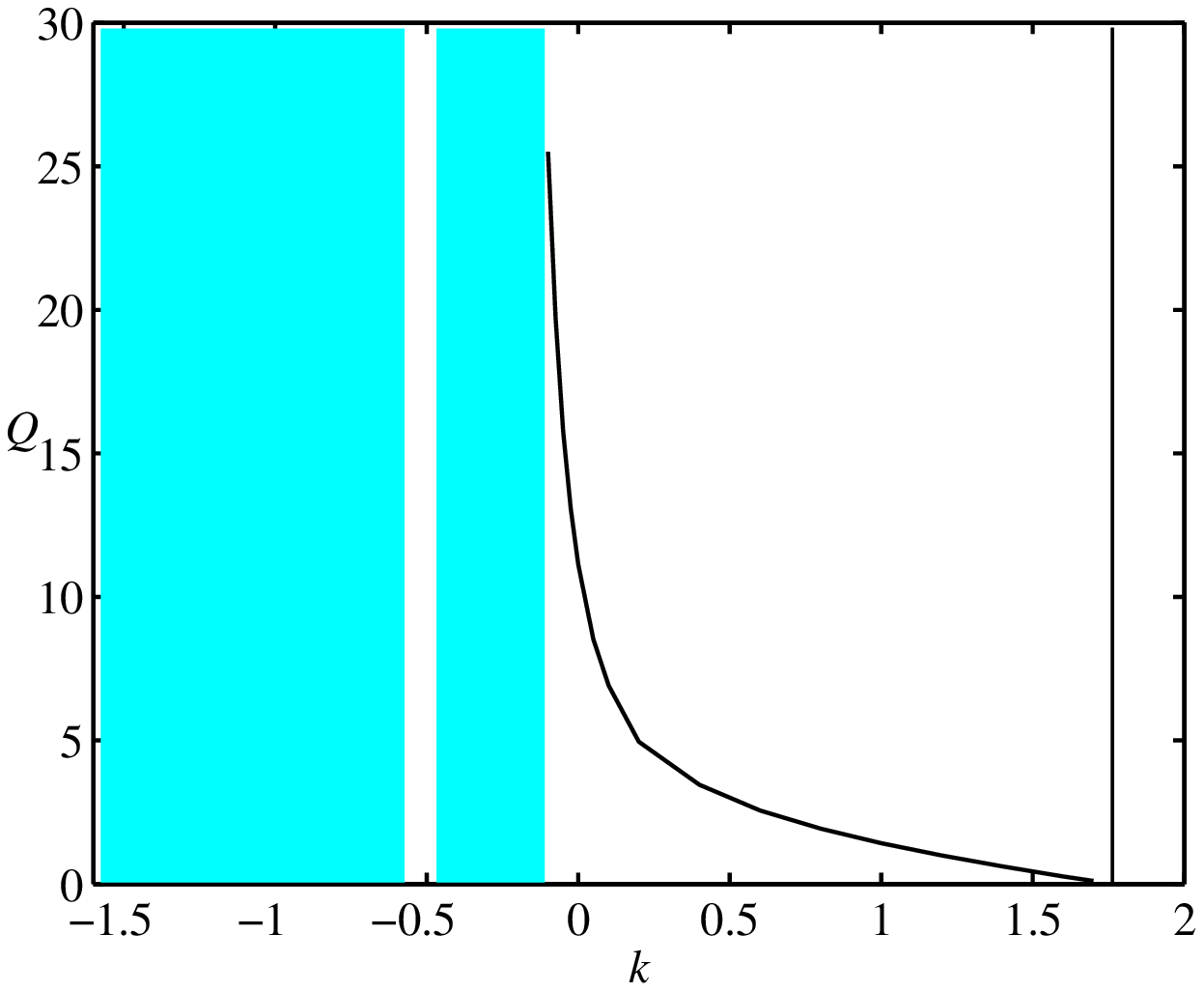}} %
\subfigure[]{\includegraphics[width=4in]{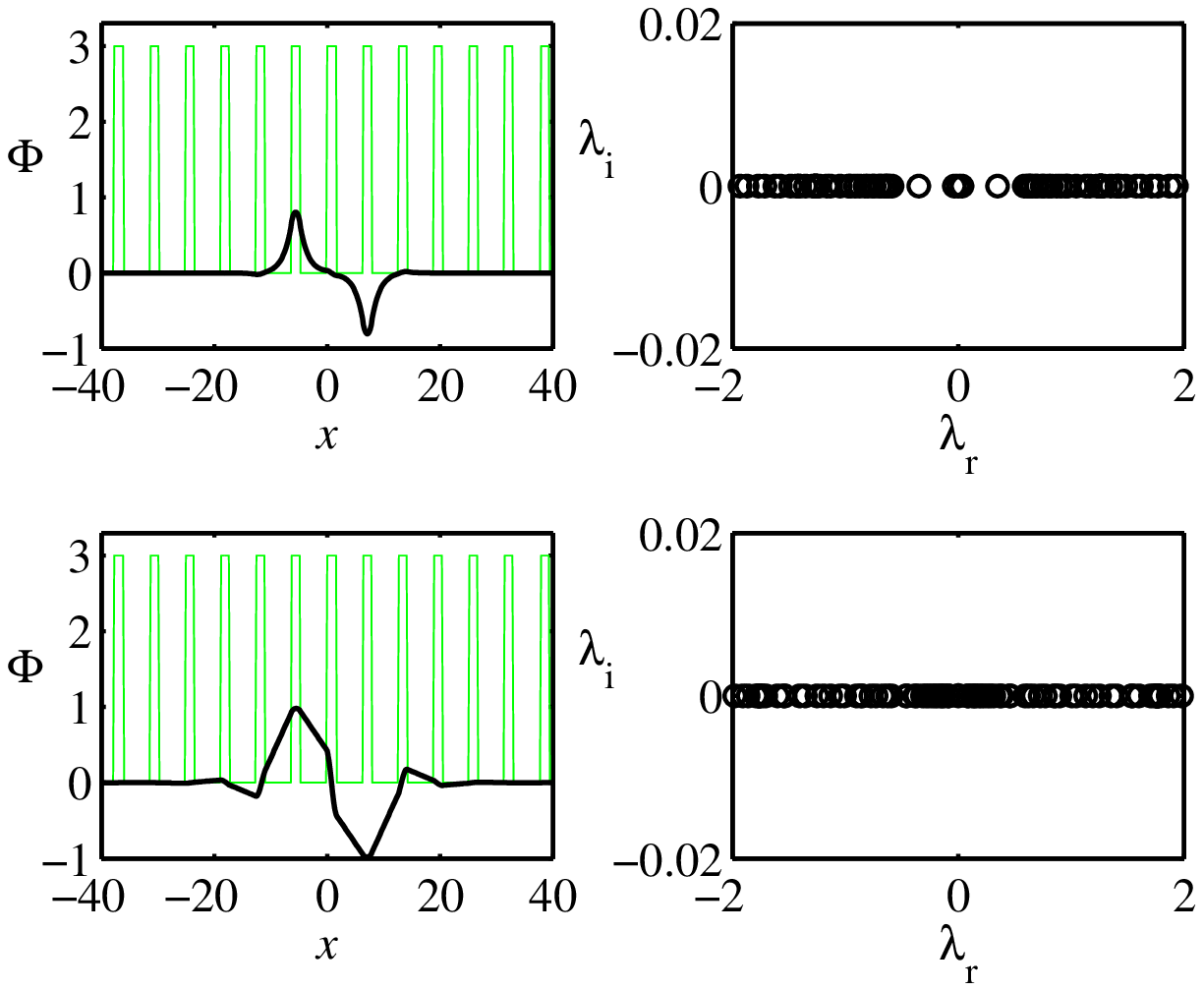}} \caption{
(a) The total power of the antisymmetric bound state versus the
propagation constant, in the first bandgap of the defocusing model
with $\mathrm{DC}=0.25$. (b) Examples of these stable
antisymmetric
bound states: (top panel) $k=0.5$ and $Q=2.96$; (bottom panel) $k=0$ and $%
Q=11.14$. }
\label{fig11}
\end{figure}

\begin{figure}[h]
\centering\subfigure[]{\includegraphics[width=4in]{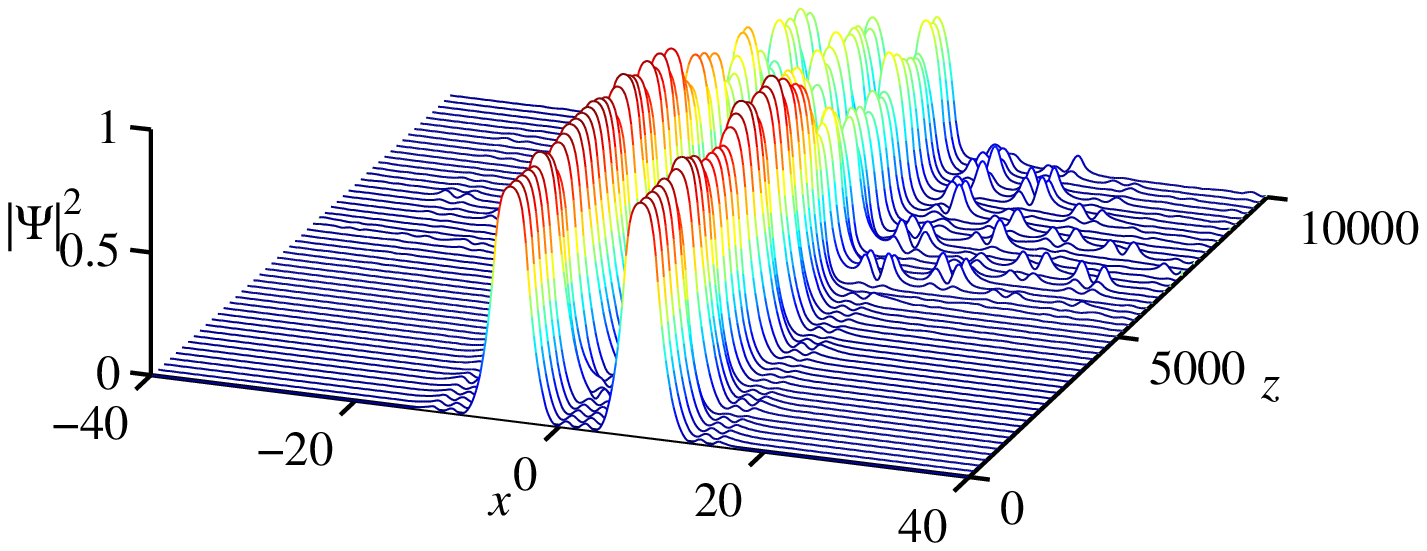}} %
\subfigure[]{\includegraphics[width=4in]{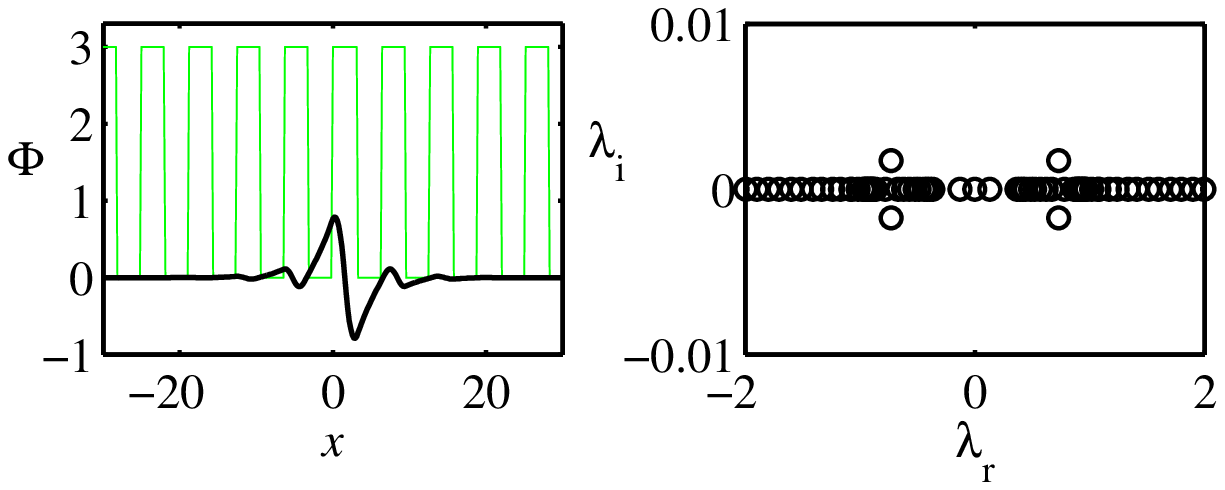}} \caption{
(a) The evolution of an unstable ``twisted" soliton (antisymmetric
bound state of two fundamental gap solitons) in the second
bandgap, for $k=0.1$ and $Q=8.17$, in the self-defocusing model with $%
\mathrm{DC}=0.5$. (b) An example of a weakly unstable subfundamental soliton
(for $k=0.1$ and $Q=2.43$) found in the second bandgap of the model with $%
\mathrm{DC}=0.50$.}
\label{fig12}
\end{figure}

In addition to the two-peak states displayed above, which may be considered
as symmetric bound states of fundamental GSs, a family of their
antisymmetric (``twisted") bound states can be readily found too. These
bound states turn out to be stable in the first finite bandgap, but unstable
in the second, both for $\mathrm{DC}=0.75$ and $0.50$. In the case of $%
\mathrm{DC}=0.25$, the second bandgap stays empty, as shown above, therefore
bound states do not exist in that bandgap either, while twisted bound states
of fundamental GSs exist and are stable in the first bandgap. These findings
are presented in Figs. \ref{fig10} and \ref{fig11}. An example of the
evolution (gradual destruction) of an unstable antisymmetric bound state in
the second bandgap is shown (for $\mathrm{DC}=0.50$) in Fig. \ref{fig12}(a).

It is known that the combination of a periodic potential and spatially
uniform SDF nonlinear term (whose power may be different from cubic \cite%
{SKA}) gives rise to an additional family of weakly unstable \textit{%
subfundamental solitons}, in the second bandgap only. The solitons were
given this name because their total power is smaller than that of the
fundamental solitons found in the same bandgap \cite{we}. Each
subfundamental soliton is an antisymmetric one, but it is \emph{not} a bound
state of fundamental GSs, as it is localized as tightly as fundamental
solitons (i.e., it is a twisted state squeezed, essentially, into a single
guiding channel). Subfundamental solitons can also be found in the second
bandgap of the present model, for $\mathrm{DC}=0.75$ and $\mathrm{DC}=0.50$
(they do not exist at $\mathrm{DC}=0.25$, as in that case the second bandgap
is empty, as shown above). For $\mathrm{DC}=0.50$, dependence $Q(k)$ for the
family of subfundamental solitons is shown by the dashed curve in Fig. \ref%
{fig10}(a), and an example of the subfundamental soliton is displayed in
Fig. \ref{fig12}(b). Similar to the ordinary model, the subfundamental
solitons in the present model are weakly unstable, and spontaneously
transform themselves into fundamental GSs belonging to the first bandgap.

\section{The self-focusing (SF)\ nonlinearity}

\begin{figure}[h]
\centering\includegraphics[width=4in]{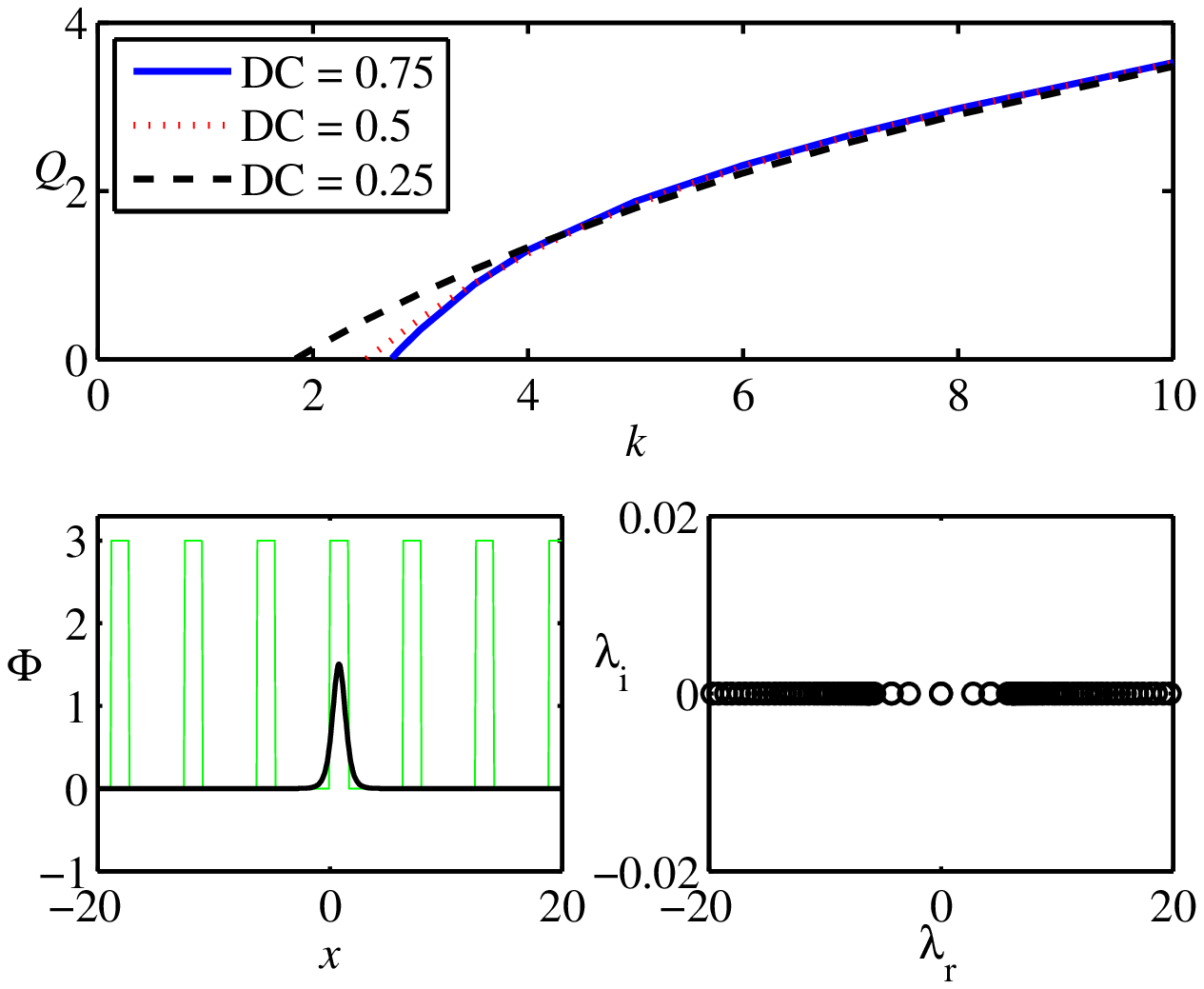} \caption{ The top
plot shows the total power versus the propagation constant for
families of solitons found in the semi-infinite gap of the
self-focusing model. The bottom plot displays a typical example of
a stable soliton, with $k=6$ and $Q=2.21$, trapped in a narrow
channel corresponding to $\mathrm{DC}=0.25$.} \label{fig13}
\end{figure}

In model with SF the sign of the nonlinear term, solitons can be readily
found in the semi-infinite gap. As might be expected, they are quite similar
to solitons previously found in the model with the KP periodic potential and
uniform nonlinearity \cite{Smerzi}. In Fig. \ref{fig13}, we display a set of
respective curves $Q\left( k\right) $ for different values of $\mathrm{DC}$,
and a typical example of a stable soliton, for $\mathrm{DC}=0.25$. The
stability of the entire soliton family complies with the prediction of the
VK criterion \cite{VK}, according to which the positive slope of the curve, $%
dQ/dk>0$, is a necessary stability condition (this criterion is irrelevant
to SDF models). As concerns the example of the soliton displayed in Fig. \ref%
{fig13}, it is worthy to note that the soliton remains trapped almost
entirely within a single waveguiding channel, even when it is narrow
(corresponding to $\mathrm{DC}=0.25$). The same feature has been observed in
all other cases.

In the parameter region investigated in the framework of this work, we have
not found multi-peak solitons in the semi-infinite gap. Solitons in finite
bandgaps of the SF model were not found either (unlike the models combining
a periodic potential and spatially uniform cubic-quintic nonlinearity, which
realizes the competition between SF and SDF terms \cite{Gisin,China}).

\section{Conclusion}

The objective of this work was to develop a systematic analysis of spatial
solitons in the model of 1D photonic crystals, built as the periodic
alternation of nonlinear waveguiding channels, of width $D$, and empty
layers between them, of width $L-D$. The main parameter of the model is the
structural ``duty cycle", $\mathrm{DC~}\equiv D/L$. We have concentrated,
chiefly, on the model with the SDF (self-defocusing) sign of the
nonlinearity in the channels, as this setting makes it possible to study new
effects due the competition between the linear trapping potential and its
effective repulsive nonlinear counterpart. Various types of solitons in this
model have been found in the first two finite bandgaps (alias
Bragg-reflection gaps) of the SDF model. The family of fundamental solitons
has been constructed too in the semi-infinite gap (alias the one accounted
for by the total internal reflection) of the system with the SF
(self-focusing) nonlinearity. In the SDF model with $\mathrm{DC}=0.75$, and\
also in the SF one, the solitons are quite similar to their counterparts in
the respective models with the uniform nonlinearity. On the other hand, the
solitons residing in the second bandgap of the SDF model with $\mathrm{DC}%
=0.50$ strongly differ from the ordinary GSs: both fundamental solitons and
their spatially symmetric bound states pass a destabilization point, with
the increase of the total power. At this point, the solitons feature a
flat-top shape, while beyond it the shape is inverted, with local maxima
emerging in empty layers. The shape inversion and destabilization of the
solitons are explained by the competition between the linear trapping
potential and its nonlinear repulsive counterpart. In the system with narrow
guiding channels, $\mathrm{DC}=0.25$, GSs are found only in the first finite
bandgap, where they are stable (both fundamental ones and their bound
states), despite the fact that they feature the inverted shape.

New properties of the fundamental solitons and their bound complexes in the
system with moderate or low values of $\mathrm{DC}$, $0.50$ and $0.25$, may
find applications to the design of all-optical data-processing schemes based
on spatial solitons in planar settings.\newline
ACKNOWLEDGMENTS\newline
The work of T. Mayteevarunyoo is supported, in a part, by a postdoctoral
fellowship from the Pikovsky-Valazzi Foundation, by the Israel Science
Foundation through the Center-of-Excellence grant No. 8006/03, and by the
Thailand Research Fund under grant No. MRG5080171.


\begin{thebibliography}{99}
\bibitem{Yabl1} E. Yablonovitch, ``Photonic band-gap structures", J. Opt.
Soc. Am. B \textbf{10}, 283-295 (1993).

\bibitem{Yabl2} E. Yablonovitch, ``Photonic crystals", J. Mod. Opt. \textbf{%
41}, 173-194 (1994).

\bibitem{Krauss} T. F. Krauss and R. M. De la Rue, ``Photonic crystals in
the optical regime -- past, present and future", Progr. Quant. Electr.
\textbf{23}, 51-96 (1999).

%
%
%

\bibitem{Nonlin} M. Soljacic and J. D. Joannopoulos, ``Enhancement of
nonlinear effects using photonic crystals", Nature Materials \textbf{3},
211-219 (2004).

\bibitem{Belgium} B. Maes, P. Bienstman, and R. Baets, ``Bloch modes and
self-localized waveguides in nonlinear photonic crystals", J. Opt. Soc. Am.
B \textbf{22}, 613-619 (2005).

\bibitem{Russell} P. S. J. Russell, ``Photonic-crystal fibers", J. Lightwave
Tech. \textbf{24}, 4729-4749 (2006).

\bibitem{RMP} E. Istrate and E. H. Sargent, ``Photonic crystal
heterostructures and interfaces", Rev. Mod. Phys. \textbf{78}, 455-481
(2006).

\bibitem{Mingaleev} K. Busch, G. von Freymann, S. Linden, S. F. Mingaleev,
L. Tkeshelashvili, and M. Wegener, ``Periodic nanostructures for photonics",
Phys. Rep. \textbf{444}, 101-202 (2007).%
%

\bibitem{John} N. Akozbek and S. John, ``Optical solitary waves in two- and
three-dimensional nonlinear photonic band-gap structures", Phys. Rev. E
\textbf{57}, 2287-2319 (1998). %
%

\bibitem{Trillo} C. Conti, S. Trillo, and G. Assanto, ``Energy localization
in photonic crystals of a purely nonlinear origin", Phys. Rev. Lett. \textbf{%
85}, 2502-2505 (2000).

\bibitem{Sammut} S. F. Mingaleev, Y. S. Kivshar, and R. A. Sammut,
``Long-range interaction and nonlinear localized modes in photonic crystal
waveguides", \ Phys. Rev. E \textbf{62}, 5777-5782 (2000).

\bibitem{KivMing} S. F. Mingaleev and Y. S. Kivshar, ``Self-trapping and
stable localized modes in nonlinear photonic crystals", Phys. Rev. Lett.
\textbf{86}, 5474-5477 (2001).

\bibitem{Sukhorukov} A. A. Sukhorukov and Y. S. Kivshar, ``Spatial optical
solitons in nonlinear photonic crystals", Phys. Rev. E \textbf{65}, 036609
(2002).

\bibitem{Sukhorukov2} A. A. Sukhorukov and Y. S. Kivshar, ``Nonlinear guided
waves and spatial solitons in a periodic layered medium", J. Opt. Soc. Am. B
\textbf{19}, 772-781 (2002).

\bibitem{ChinaPhC} P. Xie, Z. Q. Zhang, and X. D. Zhang, ``Gap solitons and
soliton trains in finite-sized two-dimensional periodic and quasiperiodic
photonic crystals", Phys. Rev. E \textbf{67}, 026607 (2003).

\bibitem{Valencia} A. Ferrando, M. Zacar\'{e}s, P. Fern\'{a}ndez de C\'{o}%
rdoba, D. Binosi, and J. A. Monsoriu, ``Spatial soliton formation in
photonic crystal fibers", Opt. Exp. \textbf{11}, 452-459 (2003).

\bibitem{Valencia2} A. Ferrando, M. Zacar\'{e}s, P. Fern\'{a}ndez de C\'{o}%
rdoba, D. Binosi, and J. A. Monsoriu, ``Vortex solitons in photonic crystal
fibers", Opt. Exp. \textbf{12}, 817-822 (2004).

\bibitem{Barcelona} Y. V. Kartashov, V. A. Vysloukh, and L. Torner,
``Soliton trains in photonic lattices", Opt. Exp. \textbf{12}, 2831-2837
(2004).

\bibitem{switch} B. Maes, P. Bienstman, and R. Baets, ``Switching in coupled
nonlinear photonic-crystal resonators", J. Opt. Soc. Am. B \textbf{22},
1778-1784 (2005).

\bibitem{Moti-theory} N. K. Efremidis, S. Sears, D. N. Christodoulides, J.
W. Fleischer, and M. Segev, ``Discrete solitons in photorefractive optically
induced photonic lattices", Phys. Rev. E \textbf{66}, 046602 (2002).

\bibitem{Moti} J. W. Fleischer, M. Segev, N. K. Efremidis, and D. N.
Christodoulides, ``Observation of two-dimensional discrete solitons in
optically induced nonlinear photonic lattices", Nature \textbf{422}, 147-150
(2003).

\bibitem{Kivshar} D. Neshev, E. Ostrovskaya, Y. Kivshar, and W. Kr\'{o}%
likowski, ``Spatial solitons in optically induced gratings", Opt. Lett.
\textbf{28}, 710-712 (2003).

\bibitem{Moti2} G. Bartal, O. Manela, O. Cohen, J. W. Fleischer, and M.
Segev, ``Observation of second-band vortex solitons in 2D photonic
lattices", Phys. Rev. Lett. \textbf{95}, 053904 (2005).

\bibitem{Sipe} N. A. R. Bhat and J. E. Sipe, ``Optical pulse propagation in
nonlinear photonic crystals", Phys. Rev. E \textbf{64}, 056604 (2001).

\bibitem{Demetri} D. N. Christodoulides and N. K. Efremidis, ``Discrete
temporal solitons along a chain of nonlinear coupled microcavities embedded
in photonic crystals", Opt. Lett. \textbf{27}, 568-570 (2002).

\bibitem{Koby} J. K. S. Poon, J. Scheuer, S. Mookherjea, G. T. Paloczi, Y.
Y. Huang, and A. Yariv, ``Matrix analysis of microring coupled-resonator
optical waveguides", Opt. Exp. \textbf{12}, 90-103 (2004).

\bibitem{Mantsyzov} B. I. Mantsyzov, I. V. Mel'nikov, and J. Stewart
Aitchison, ``Controlling light by light in a one-dimensional resonant
photonic crystal", Phys. Rev. E \textbf{69}, 055602(R) (2004).

\bibitem{Herrmann} A. V. Husakou and J. Herrmann, ``Supercontinuum
generation of higher-order solitons by fission in photonic crystal fibers",
Phys. Rev. Lett. \textbf{87}, 203901 (2001).

\bibitem{Herrmann-experiment} J. Herrmann, U. Griebner, N. Zhavoronkov, A.
Husakou, D. Nickel, J. C. Knight, W. J. Wadsworth, P. S. J. Russell, and G.
Korn, ``Experimental evidence for supercontinuum generation by fission of
higher-order solitons in photonic fibers", Phys. Rev. Lett. \textbf{88},
173901 (2002).

\bibitem{Russel-soliton} F. Luan, J. C. Knight, P. St. J. Russell, S.
Campbell, D. Xiao, D. T. Reid, B. J. Mangan, D. P. Williams, and P. J.
Roberts, ``Femtosecond soliton pulse delivery at 800nm wavelength in
hollow-core photonic bandgap fibers", Opt. Exp. \textbf{12}, 835-840 (2004).

\bibitem{Skryabin} A. Efimov, A. V. Yulin, D. V. Skryabin, J. C. Knight, N.
Joly, F. G. Omenetto, A. J. Taylor, and P. Russell, ``Interaction of an
optical soliton with a dispersive wave", Phys. Rev. Lett. \textbf{95},
213902 (2005).%
%
%

\bibitem{Bang} J. F. Corney and O. Bang, ``Solitons in quadratic nonlinear
photonic crystals", Phys. Rev. E Volume: \textbf{64}, 047601 (2001).

\bibitem{KP} C. Kittel, \textit{Introduction to Solid State Physics} (Wiley:
New York, 1995).

\bibitem{Smerzi} W. Li and A. Smerzi, Phys. Rev. E 70, 016605 (2004).

\bibitem{Carr} B. T. Seaman, L. D. Carr, and M. J. Holland, ``Nonlinear band
structure in Bose-Einstein condensates: Nonlinear Schr\"{o}dinger equation
with a Kronig-Penney potential", Phys. Rev. A \textbf{71}, 033622 (2005).

\bibitem{Gisin} I. M. Merhasin, B. V. Gisin, R. Driben, and B. A. Malomed,
``Finite-band solitons in the Kronig-Penney model with the cubic-quintic
nonlinearity", Phys. Rev. E \textbf{71}, 016613 (2005).

\bibitem{China} J. Wang, F. Ye, L. Dong, T. Cai, and Y.-P. Li, ``Lattice
solitons supported by competing cubic--quintic nonlinearity", Phys. Lett. A
\textbf{339}, 74 (2005).

\bibitem{Salerno} G. L. Alfimov, V. V. Konotop, and M. Salerno, Europhys.
Lett., ``Matter solitons in Bose-Einstein condensates with optical
lattices", Europhys. Lett. \textbf{58}, 7-13 (2002).

\bibitem{Salerno2} B. B. Baizakov, V. V. Konotop, and M. Salerno, ``Regular
spatial structures in arrays of Bose--Einstein condensates induced by
modulational instability", J. Phys. B: At. Mol. Opt. Phys. \textbf{35},
5105--5119 (2002).

\bibitem{Louis} P. J. Y. Louis, E. A. Ostrovskaya, C. M. Savage, and Y. S.
Kivshar, ``Bose-Einstein condensates in optical lattices: Band-gap structure
and solitons", Phys. Rev. A \textbf{67}, 013602 (2003).

\bibitem{Sakaguchi} H. Sakaguchi and B. A. Malomed, ``Matter-wave solitons
in nonlinear optical lattices", Phys. Rev. E \textbf{72}, 046610 (2005).

\bibitem{Fibich} Y. Sivan, G. Fibich, and M. I. Weinstein, ``Waves in
nonlinear lattices: Ultrashort optical pulses and Bose-Einstein
condensates", Phys. Rev. Lett. \textbf{97}, 193902 (2006).

\bibitem{Fatkhulla} F. Abdullaev, A. Abdumalikov, and R. Galimzyanov, ``Gap
solitons in Bose--Einstein condensates in linear and nonlinear optical
lattices", Phys. Lett. A 367, 149-155 (2007).

\bibitem{lin-nonlin-competition} Z. Rapti, P. G. Kevrekidis, V. V. Konotop
and C. K. R. T. Jones, ``Solitary waves under the competition of linear and
nonlinear periodic potentials", \ J. Phys. A: Math. Theor. \textbf{40},
14151-14163 (2007).

\bibitem{we} T. Mayteevarunyoo and B. A. Malomed, ``Stability limits for gap
solitons in a Bose-Einstein condensate trapped in a time-modulated optical
lattice", Phys. Rev. A \textbf{74}, 033616 (2006).

\bibitem{Wang} B. A. Malomed, Z. H. Wang, P. L. Chu, and G. D. Peng,
``Multichannel switchable system for spatial solitons", J. Opt. Soc. Am. B
\textbf{16}, 1197-1203 (1999).

\bibitem{VK} N. G. Vakhitov and A. A. Kolokolov, ``Stationary solutions of
the wave equation in a medium with nonlinearity saturation", Izv. Vyssh.
Uchebn. Zaved., Radiofiz. \textbf{16}, 10120 (1973) [in Russian; English
translation: Radiophys. Quantum. Electron. \textbf{16}, 783 (1973)]; see
also L. Berg\'{e}, ``Wave collapse in physics: principles and applications
to light and plasma waves", Phys. Rep. \textbf{303}, 259-370 (1998).

\bibitem{spectrum} B. Deconinck, F. Kiyak, J. D. Carter, and J. N. Kutz,
``SpectrUW: A laboratory for the numerical exploration of spectra of linear
operators, ``Math. Comput. Simul. \textbf{74}, 370-378 (2007).

\bibitem{SKA} S. Adhikari and B. A. Malomed, \textquotedblleft Tightly bound
gap solitons in a Fermi gas", Europhys. Lett. \textbf{79}, 50003 (2007).
\end{thebibliography}
\end{document}